\pdfoutput=1

\documentclass[11pt]{article}

\usepackage[final]{acl}

\usepackage{times}
\usepackage{latexsym}

\usepackage[T1]{fontenc}

\usepackage[utf8]{inputenc}

\usepackage{microtype}
\usepackage{colortbl}
\usepackage{graphicx}
\usepackage{multirow}
\usepackage{amsmath}
\usepackage{enumitem}
\usepackage{subcaption}
\usepackage{xcolor}
\usepackage{amssymb}

\title{In-context Contrastive Learning for Event Causality Identification}

\author{Chao Liang \textsuperscript{1} \hspace{1cm} Wei Xiang \textsuperscript{2} \hspace{1cm} Bang Wang \textsuperscript{1} \thanks{\quad Corresponding author: Bang Wang} \\
	\textsuperscript{1} School of Electronic Information and Communications, \\
	Huazhong University of Science and Technology, Wuhan, China \\
	\texttt{\{liangchao111, wangbang\}@hust.edu.cn} \\
	\textsuperscript{2} Faculty of Artificial Intelligence in Education, \\
	Central China Normal University, Wuhan, China. \\
	\texttt{xiangwei@ccnu.edu.cn}}

\begin{document}
\maketitle
\begin{abstract}
Event Causality Identification (ECI) aims at determining the existence of a causal relation between two events. Although recent prompt learning-based approaches have shown promising improvements on the ECI task, their performance are often subject to the delicate design of multiple prompts and the positive correlations between the main task and derivate tasks. The in-context learning paradigm provides explicit guidance for label prediction in the prompt learning paradigm, alleviating its reliance on complex prompts and derivative tasks. However, it does not distinguish between positive and negative demonstrations for analogy learning. Motivated from such considerations, this paper proposes an \textbf{I}n-\textbf{C}ontext \textbf{C}ontrastive \textbf{L}earning (ICCL) model that utilizes contrastive learning to enhance the effectiveness of both positive and negative demonstrations. Additionally, we apply contrastive learning to event pairs to better facilitate event causality identification. Our ICCL is evaluated on the widely used corpora, including the EventStoryLine and Causal-TimeBank, and results show significant performance improvements over the state-of-the-art algorithms. \footnote{\space We release the code at: \url{https://github.com/ChaoLiang-HUST/ICCL}.}

\end{abstract}

\section{Introduction}
\begin{figure}[t]
	\centering
	\includegraphics[width=\columnwidth , trim=25 510 235 10, clip]{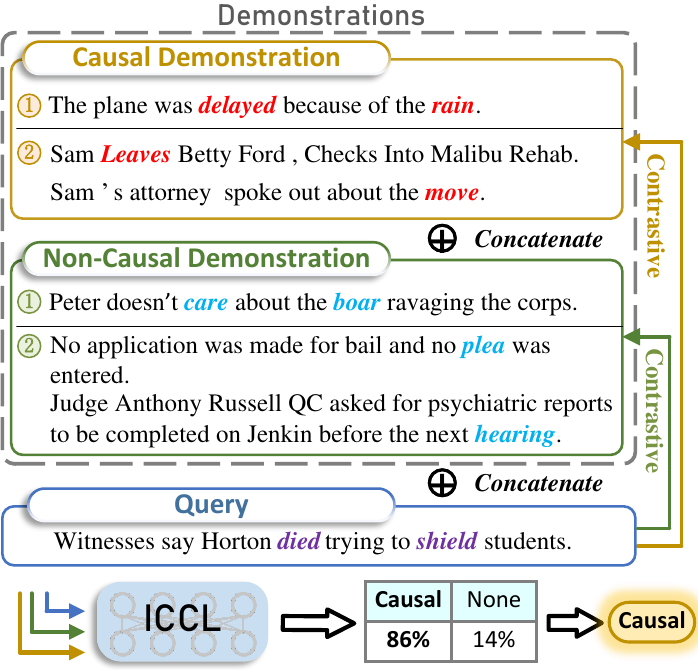}
	\caption{Illustration of our motivation. The event pairs are highlighted in different colors.}
	\label{fig:demos}
	\vspace{-10pt}
\end{figure}

Event Causality Identification (ECI) is to detect whether there exists a causal relation between two event mentions in a document. It is of great importance for many Natural Language Processing (NLP) applications, such as question answer \citep{Breja:et.al:COLINS:2020}, machine reading comprehension \citep{Berant:et.al:EMNLP:2014}, and etc. Furthermore, It also has many practical applications in real-world scenarios, such as event prediction \citep{Preethi:et.al:science:2015,Radinsky:et.al:WWW:2012} and strategy optimization \citep{Balgi:et.al:AAAI:2022}.
Fig.~\ref{fig:demos} illustrates an event causality example from the Event StoryLine Corpus (ESC). We concatenated two causal demonstrations and two non-causal demonstrations before the query to be predicted, and enhanced the analogy between the query and demonstrations through contrastive. Ultimately, our ICCL model determined the causality between the two events, "\textit{died}" and "\textit{shield}", in the query.

\par
Some graph-based methods have been proposed for the ECI task \citep{Zhao:et.al:Sciences:2021, Phu:et.al:NAACL:2021, Pu:et.al:ACL:2023}, which apply a graph structure to represent events and their potential relations. For example, \citet{Zhao:et.al:Sciences:2021} initialize event nodes' embeddings using a document-level encoder and employ a graph inference mechanism to update their embeddings. \citet{Pu:et.al:ACL:2023} incorporate causal label information and event pair interaction information to enhance the representation learning for event nodes in the graph. These methods follow the traditional representation learning for classification yet on a graph structure.

\par
Recently the prompt learning paradigm \citep{Liu:ey.al:ACM:2023} has shown its great successes in many NLP tasks, as it can well leverage the potentials of a pre-trained language model (PLM). Some researchers have applied the prompt learning for the ECI task \citep{Liu:et.al:IJCAI:2021,Shen:et.al:COLING:2022}. For example, the DPJL model \citep{Shen:et.al:COLING:2022} designs a main cloze task but also designs two derivative prompt tasks. Although the DPJL has achieved new state-of-the-art performance, it involves the delicate design of multiple prompts and relies on the positive correlations between the main task and derivative tasks.

\par
The in-context learning paradigm ~\cite{Dong:et.al:arXiv:2022} includes some demonstrations with their ground-truth labels into the query prompt to learn some patterns hidden in demostrations when making its prediction. However, it does not distinguish between positive and negative demonstrations for analogy. Motivated from such considerations, we propose to use contrastive learning on the in-context demonstrations to enhance the effectiveness of analogy, as illustrated in Fig.~\ref{fig:demos}. Besides, we also argue that the semantic of event mentions are the most important for the causal relation identification between them. As such we apply contrastive learning to the representation of event mentions in in-context demonstrations, so as to distinguishing the semantic between causal and non-causal event pairs and facilitating event causality predictions.

\par
In this paper, we propose an \textbf{I}n-\textbf{C}ontext \textbf{C}ontrastive \textbf{L}earning (\textbf{ICCL}) model for the ECI task.
The ICCL model contains three modules. The prompt learning module reformulates an input event pair and some retrieved demonstrations into a prompt template, as the input for PLM encoding. The in-context contrastive module optimizes the representation of event mention by simultaneously maximizing its agreement with positive demonstrations and minimizing with negative ones, via a contrastive loss. The causality prediction module predicts answer word to identify causal relations. Experiments are conducted on the widely used EventStoryLine and Causal-TimeBank corpora, and results have shown that our ICCL achieves the new state-of-the-art performance for the ECI task.

\begin{figure*}[h]
	\centering
	\includegraphics[width=1\textwidth , trim=30 180 50 70, clip]{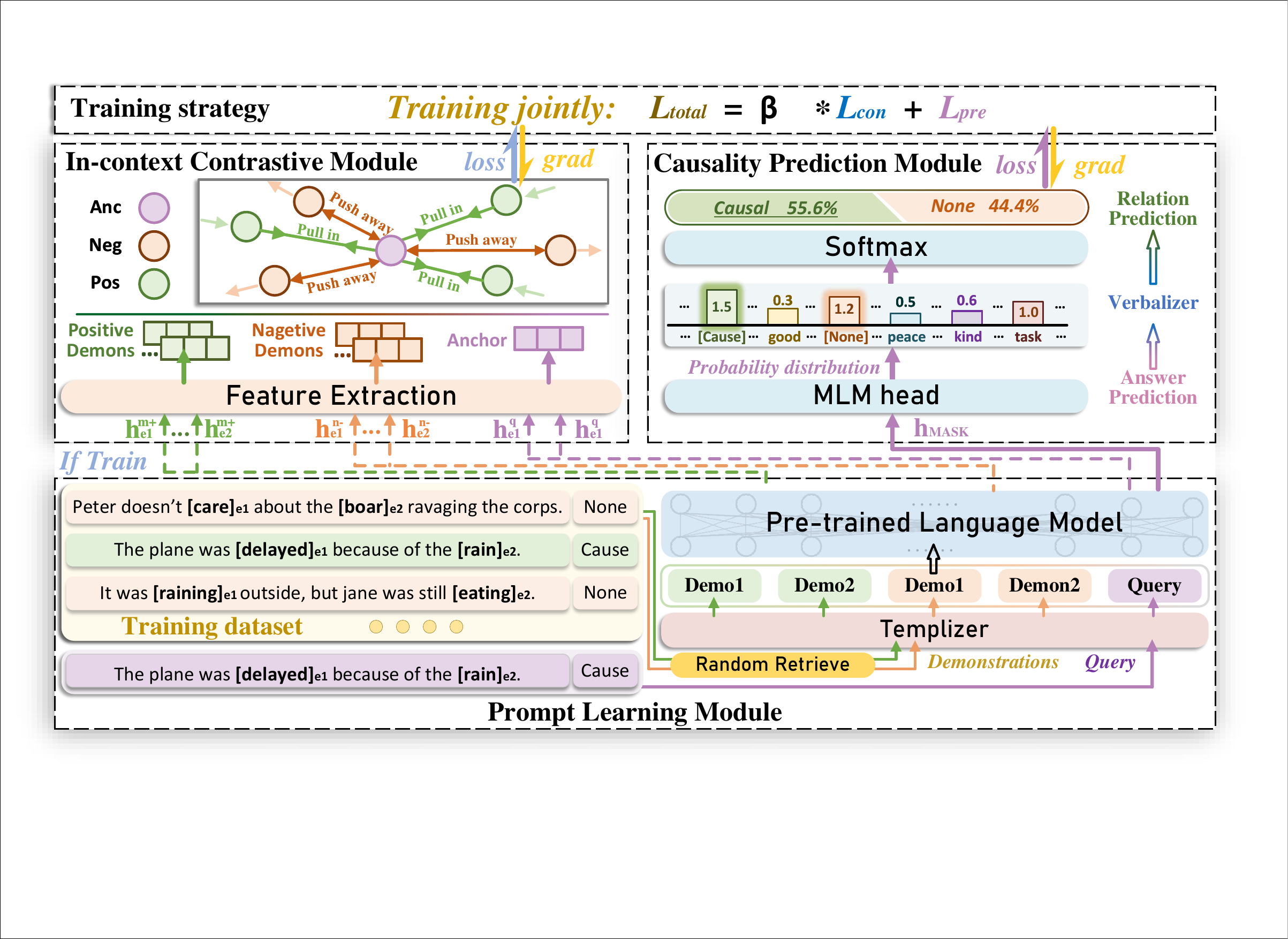}
	\caption{Illustration of our ICCL framwork.}
	\label{fig:model}
\end{figure*}

\section{Related work}

\subsection{Event Causality Identification}
Event Causality Identification (ECI) is an essential task in information extraction, attracting significant attention due to its wide range of potential applications. Early methods mainly relied on designing task-oriented neural network models \citep{Liu:et.al:IJCAI:2021, Zuo:et.al:arXiv:2021}. For example, \citet{Liu:et.al:IJCAI:2021} improve the capability of their neural model to identify previously unseen causal relations by incorporating event-agnostic and context-specific patterns derived from the ConceptNet \citep{Speer:et.al:AAAI:2017}. With further exploration of graph structures and the emergence of large-scale PLMs, recent studies have increasingly adopted graph-based and prompt-based learning approaches to address the ECI task.

\par
Graph-based approaches usually model the ECI task as a node classification problem, employing graph neural networks to learn event node representations based on contextual semantics at the document level \citep{Phu:et.al:NAACL:2021, Cao:et.al:ACL:2021, Fan:et.al:SIGIR:2022}. For example, \citet{Fan:et.al:SIGIR:2022} establish explicit connections between events, mentions and contexts to construct a co-occurrence graph for node representation learning and causal relation identification. In addition to node classification, some studies approach the ECI task as a graph-based edge prediction problem \citep{Zhao:et.al:Sciences:2021, Chen:et.al:arXiv:2022}. For example, \citet{Zhao:et.al:Sciences:2021} initialize event node embeddings using a document-level encoder based on a PLM and employ a graph inference mechanism to predict causal edges through graph updating.

\subsection{Prompt-based Causality Identification}
Recently, with the help of large-scale PLMs, such as the BERT \citep{Devlin:et.al:arXiv:2018}, RoBERTa \citep{Liu:et.al:arXiv:2019} and etc, prompt learning has emerged as a new paradigm for various NLP tasks \citep{Xiang:et.al:COLING:2022, Ding:et.al:arXiv:2021}. It converts downstream tasks into the similar form as pre-training task, which aligns objectives between the two stages. This alignment helps bridging the gap between PLM and task and can directly enhance the performance of a downstream task. Moreover, researchers have also devised appropriate prompts to reframe ECI task as a cloze task \citep{Shen:et.al:COLING:2022, Liu:et.al:IJCAI:2021}. For example, \citet{Shen:et.al:COLING:2022} propose a derivative prompt joint learning model that leverages potential causal knowledge within PLMs based on the causal cue words detection. \citet{Liu:et.al:IJCAI:2021} use an event mention masking generalization mechanism to encode some event causality patterns for causal relation reasoning. Although prompt-based methods are constrained by complex prompts and derivate tasks, these prompt-based models effectively leverage the implicit knowledge of PLMs to address the ECI task.

\section{Method}
Fig.~\ref{fig:model} illustrates our ICCL model, including the prompt learning module, the in-context contrastive module, and the causality prediction module.

\subsection{Task Formulation}
We apply the prompt learning paradigm to transform the ECI task into a causal relation cloze task, utilizing a PLM to predict answer words for causal relation identification. As the event mentions are annotated by a few words in a sentence, we use the event mentions $E_1$ and $E_2$ of an event pair as well as their raw sentences $S_1$ and $S_2$, as the input $x = \{ E_1, E_2, S_1, S_2\}$, where $E_1 \in S_1$ and $E_2 \in S_2$.
The virtual answer words \texttt{<causal>} and \texttt{<none>} indicating whether there is a causal relation between the input event pair, are used as the output $y \in \{\texttt{<causal>}, \texttt{<none>}\}$.
We note that in cases where $E_1$ and $E_2$ originate from the same sentence, $S_1$ and $S_2$ refer to the same sentence.

\subsection{Prompt Learning Module}
As illustrated in the bottom of Fig.~\ref{fig:model}, we first reformulate each input instance $x = \{ E_1, E_2, S_1, S_2\}$ into a kind of in-context prompt template $T(x)$, as the input of a PLM for encoding. The in-context prompt input contains a query instance and $K$ retrieved demonstrations. The query instance is the input event instance, denoted as $q=\{E_1^q, E_2^q, S_1^q, S_2^q\}$, with the causal relation between two events to be identified.
The demonstrations are retrieved from the training dataset, consisting of an event mention pair and their raw sentences, as well as the relation label between them, denoted as $d_k=\{E_1^k, E_2^k, S_1^k, S_2^k, y^k\}$. We randomly select $M$ demonstrations labeled with \texttt{<causal>} relation and $N$ demonstrations labeled with \texttt{<none>} relation, denoted as $d_m^+$ and $d_n^-$, respectively.

\par
We design a prediction prompt template $T_p(q)$ for the query instance $q$ and an analogy prompt template $T_a(d_k)$ for its retrieved demonstrations $d_k$, respectively. Both of them are constructed by concatenating the raw sentences with a simple cloze template, as follows:
\begin{align*}
T_p(q) = & \; S_1^q + S_2^q \; + \\
&  \mathtt{[start]} + E_1^q + \mathtt{[MASK]} + E_2^q + \mathtt{[end]}. \\
T_a(d_k) = & \; S_1^k + S_2^k \; + \\
&  \mathtt{[start]} + E_1^k + y^k + E_2^k + \mathtt{[end]}.
\end{align*}
where $E_1^q, E_2^q, S_1^q, S_1^q$ are the two event mentions and their raw sentences, and the PLM-specific token $\mathtt{[start]}$ and $\mathtt{[end]}$ are used to indicate the beginning and ending of the cloze template.
For prediction prompt template $T_p(q)$, a PLM-specific token $\mathtt{[MASK]}$ is inserted between two event mentions for relation prediction; For analogy prompt template $T_a(d_k)$, it is replaced by the virtual word of the relation label $y^k$ for each demostrations, i.e. \texttt{<causal>} or \texttt{<none>}.

\par
The in-context prompt template $T(x)$ is constructed by concatenating the prediction prompt tempalte $T_p(q)$ and some analogy prompt templates $T_a(d_k)$ of its retrieved demonstrations, as follows:
\begin{align*}
	T(x) = \mathtt{[CLS]} + T_a(d_1^+) \; \mathtt{[SEP]} \dots T_a(d_M^+) \; \mathtt{[SEP]} \\
	 + \; T_a(d_1^-) \; \mathtt{[SEP]} \dots T_a(d_N^-) \; \mathtt{[SEP]} + T_p(q) \; \mathtt{[SEP]}.
\end{align*}
where the PLM-specific token $\mathtt{[CLS]}$ and $\mathtt{[SEP]}$ are used to indicate the beginning and ending of an input, and some  $\mathtt{[SEP]}$ tokens are used as separators between the query and those demonstrations.
Note that, the causal demonstrations $d_m^+$ are positioned before the none causal demonstrations $d_n^-$.
We provide a specific example of in-context prompt template input in Appendix~\ref{sec:input}.

\par
After the PLM encoding, we obtain a hidden state $\mathbf{h} \in \mathbb{R}^d$ for each input tokens, where $d$ is the dimension of hidden states. We denote the hidden state of input $\mathtt{[MASK]}$ token as ${\mathbf{h}_{mask}}$ for causality prediction.
The hidden states of input event pair in query instance, retrieved causal and none-causal demonstrations are denoted as $[{\mathbf{h}_{e_1}^q}, {\mathbf{h}_{e_2}^q}]$, $[{\mathbf{h}_{e_1}^{m^+}}, {\mathbf{h}_{e_2}^{m^+}}]$ and $[{\mathbf{h}_{e_1}^{n^-}}, {\mathbf{h}_{e_2}^{n^-}}]$, respectively, which are next used for in-context contrastive learning.

\subsection{In-context Contrastive Module}
The in-context contrastive module optimizes the representation of event mention by simultaneously maximizing its agreement with positive demonstration samples and minimizing with negative ones, via a contrastive loss. In the training phase, we use the input query instance as an anchor. The retrieved demonstrations with the same relation label as the query are positive samples, while those with different relation label are negative samples.
We assume that the query's label is \texttt{<causal>}, so the causal demonstrations $d_m^+$ being treated as positives, and non-causal ones $d_n^-$ as negatives.

\par
Motivated by the fact that the offsets of pre-trained word embeddings can model the relationship between them~\citep{Mikolov:et.al:arXiv:2013, Pennington:et.al:EMNLP:2014, Chen:et.al:arXiv:2016}, such as  $\textbf{h}_{king} - \textbf{h}_{man} \approx \textbf{h}_{queen}-\textbf{h}_{woman}$. We use the offsets between event mentions' hidden states to represent their relation for contrastive learning, as follows:
\begin{align}
	\mathbf{z}^q &= \mathbf{h}_{e_1}^q - \mathbf{h}_{e_2}^q \label{eq:1}, \\
	\mathbf{z}^+_m &= \mathbf{h}_{e_1}^{m^+} - \mathbf{h}_{e_2}^{m^+} \label{eq:2}, \\
	\mathbf{z}^-_n &= \mathbf{h}_{e_1}^{n^-} - \mathbf{h}_{e_2}^{n^-} \label{eq:3},
\end{align}
where $\mathbf{z}^q, \mathbf{z}_m^+, \mathbf{z}_n^-$ are the relation vector of event pair in query instance, positive and negative demonstrations, respectively.

\par
We adpot supervised constrastive learning on the relation vector of event pair for its representation optimization~\citep{Khosla:et.al:NeurIPS:2020}. Specifically, it pulls together the anchor towards positive samples in embedding space, while simultaneously pushing it apart from negative samples.
The supervised contrastive loss is computed as follows:
\begin{align}
	L_{con} = -\log\sum_{m=1}^{M}{\dfrac{\exp({\text{sim}(\mathbf{z}^q, \mathbf{z}^+_m)}/{\tau})}		{\sum\limits_{d \in \mathcal{D}}{\exp({\text{sim}(\mathbf{z}^q, d)}/{\tau})}}},	\label{eq:4}
\end{align}
where ${\mathcal{D} = \{\mathbf{z}^+_m\}_{m=1}^{M} \cup \{\mathbf{z}^-_n\}_{n=1}^{N}}$, $M$ and $N$ represent the number of positive and negative demonstrations, respectively.

\subsection{Causality Prediction Module}
The causality prediction module uses the $\mathtt{[MASK]}$ token of input query instance to predict an answer word for causal relation identification.
Specifically, we input the hidden state $\mathbf{h}_{mask}$ into the masked language model classifier, and estimate the probability of each word in its vocabulary dictionary $\mathcal{V}$ for the $\mathtt{[MASK]}$ token, as follows:
\begin{align}
	P(\mathtt{[MASK]} = v \in \mathcal{V} \; | \; T(x)), \label{eq:5}
\end{align}

\par
We add two virtual words into PLM's vocabulary dictionary as the answer space, viz. \texttt{<causal>} and \texttt{<none>} , to indicate whether a causal relation exists or not.
Then a softmax layer is applied on the prediction scores of the two virtual answer words to normalize them into probabilities:
\begin{align}
	P_i(v_i \in \mathcal{V}_a | T(x)) = \dfrac{\exp(p_{v_i})}{\sum_{j=1}^{n}\exp(p_{v_j})}, \label{eq:6}
\end{align}
where $\mathcal{V}_a = \{ \texttt{<causal>, \texttt{<none>}\}}$.
\par

In the training phase, we tune parameters of PLM and MLM classifier based on in-context prompt and newly added vitual words. We adopt the cross entropy loss as the loss function:
\begin{align}
	L_{pre} = -\frac{1}{L} \sum\limits_{l=1}^{L} \mathbf{y}^{(l)} \log(\mathbf{\hat{y}}^{(l)}) + \lambda \Vert \theta \Vert^2,  \label{eq:7}
\end{align}
where $\mathbf{y}^{(l)}$ and $\mathbf{\hat{y}}^{(l)}$ are answer label and predicted label of the $l$-th training instance respectively. $\lambda$ and $\theta$ are the regularization hyper-parameters. We use the AdamW optimizer \citep{Loshchilov:et.al:arXiv:2017} with $L2$ regularization for model training.

\subsection{Training strategy}
We jointly train the in-context contrastive module and the causality prediction module.
The loss function of our ICCL model is optimized as follows:
\begin{align}
	L_{total} = L_{pre} + \beta * L_{con}, \label{eq:8}
\end{align}
where $\mathtt{\beta}$ is the weight coefficient to balance the importance of contrastive loss and prediction loss.
We conduct some experiments to explore the impact of different $\mathtt{\beta}$ values on model performance. The experimental results and analysis are presented in Appendix~\ref{sec:beta}.

\section{Experiment Setting}
\subsection{Datasets}
Our experiments are conducted on two widely used datasets for the ECI task: EventStory-Line 0.9 Corpus (ESC) \citep{Caselli:et.al:ACL:2017} and Causal-TimeBank Corpus (CTB) \citep{Mirza:et.al:COLING:2014}.
\par

\textbf{EventStoryLine} contains 22 topics and 258 documents collected from various news websites. In total, there are 5,334 event mentions in ECS dataset. Among them, 5,625 event pairs are annotated with causal relations. Specifically, 1,770 causal relations are intra-sentence causalities, while 3,855 ones are cross-sentence causalities.
Following the standard data splitting~\citet{Gao:et.al:NAACL:2019}, we use the last two topics as the development set, and conduct 5-fold cross-validation on the remaining 20 topics. The average results of precision (P), recall (R), and F1 score are adopted as performance metrics.

\par
\textbf{Causal-TimeBank} comprises 184 documents sourced from English news articles, with a total of 7,608 annotated event pairs. Among them, 318 are annotated with causal relations. Specifically, 300 causal relations are intra-sentence causalities, while only 18 ones are cross-sentence causalities. Following the standard data splitting~\cite{Liu:et.al:arXiv:2021}, we employ a 10-fold cross-validation and the average results of precision (P), recall (R), and F1 score are adopted as performance metrics. Following \citet{Phu:et.al:NAACL:2021}, we only conduct intra-sentence event causality identification experiments on CTB, as the number of cross-sentence event causal pairs is quite small.

\begin{table*}[ht]
	\centering
	\resizebox{\textwidth}{!}{
		\renewcommand\arraystretch{1.2}
		
		\begin{tabular}{c|c|ccc|ccc|ccc|ccc}
			\hline
			\multirow{3}{*}{\textbf{Model}} & \multirow{3}{*}{\textbf{PLM}} & \multicolumn{9}{c|}{\textbf{EventStoryLine}}& \multicolumn{3}{c}{\textbf{Causal-TimeBank}}\\
			\cline{3-14}
			& & \multicolumn{3}{c|}{Intra} & \multicolumn{3}{c|}{Cross} & \multicolumn{3}{c|}{Intra and Cross}& \multicolumn{3}{c}{Intra}\\
			\cline{3-14}
			& & P(\%) & R(\%) & F1(\%) & P(\%) & R(\%) & F1(\%) & P(\%) & R(\%) & F1(\%)& P(\%) & R(\%) & F1(\%)\\
			\hline
			ILP \citep{Gao:et.al:NAACL:2019} & - & 38.8 & 52.4 & 44.6 & 35.1 & 48.2 & 40.6 & 36.2 & 49.5 & 41.9 & - & - & - \\
			LearnDA \citep{Zuo-b:et.al:arXiv:2021} & BERT & 42.2 & 69.8 & 52.6 & - & - & - & - & - & - & 41.9 & \cellcolor[rgb]{1, 0.9, 0.8} 68.0 & 51.9 \\
			RichGCN \citep{Phu:et.al:NAACL:2021} & BERT & 49.2 & 63.0 & 55.2 & 39.2 & 45.7 & 42.2 & 42.6 & 51.3 & 46.6 & 39.7 & 56.5 & 46.7 \\
			DPJL \citep{Shen:et.al:COLING:2022} & RoBERTa & 65.3 & \cellcolor[rgb]{1, 0.9, 0.8} 70.8 & \cellcolor[rgb]{1, 0.9, 0.8} 67.9 & - & - & - & - & - & - & 63.6 & 66.7 & 64.6 \\
			GESI \citep{Fan:et.al:SIGIR:2022} & BERT & - & - & 50.3 & - & - & 49.3 & - & - & 49.4& - & - & - \\
			ERGO \citep{Chen:et.al:arXiv:2022} & Longformer & 57.5 & 72.0 & 63.9 & 51.6 & 43.3 & 47.1 & 48.6 & 53.4 & 50.9& 62.1 & 61.3 & 61.7 \\
			SemSln \citep{Hu:et.al:arXiv:2023} & BERT & 64.2 & 65.7 & 64.9 & - & - & - & - & - & - & 52.3 & 65.8 & 58.3 \\
			\hline
			\multirow{4}{*}{\textbf{ICCL}}
			 & BERT & 64.9 & 69.6 & 67.1 & 56.3 & \cellcolor[rgb]{1, 0.9, 0.8} 58.4 & 57.2 & 59.0 & 61.9 & 60.4& 60.5 & 58.4 & 59.1\\
			
			 & ERNIE & 66.8 & 68.5 & 67.5 & \cellcolor[rgb]{1, 0.8, 0.6} 63.7 & 56.2 & 59.5 & \cellcolor[rgb]{1, 0.8, 0.6} 64.8 & 60.0 & 62.1& \cellcolor[rgb]{1, 0.9, 0.8} 64.8 & 66.0 & 64.7 \\
			
			 & DeBERTa & \cellcolor[rgb]{1, 0.8, 0.6} 67.6 & \cellcolor[rgb]{1, 0.8, 0.6} 73.7 & \cellcolor[rgb]{1, 0.8, 0.6} 70.4 & \cellcolor[rgb]{1, 0.9, 0.8} 61.8 & \cellcolor[rgb]{1, 0.9, 0.8} 58.4 & \cellcolor[rgb]{1, 0.9, 0.8} 59.9 & 61.7 & \cellcolor[rgb]{1, 0.9, 0.8} 63.2 & \cellcolor[rgb]{1, 0.9, 0.8} 63.3& \cellcolor[rgb]{1, 0.8, 0.6} 66.7 & 64.4 & \cellcolor[rgb]{1, 0.9, 0.8} 64.9 \\
			
			 & RoBERTa & \cellcolor[rgb]{1, 0.9, 0.8} 67.5 & \cellcolor[rgb]{1, 0.8, 0.6} 73.7 & \cellcolor[rgb]{1, 0.8, 0.6} 70.4 & 60.3 & \cellcolor[rgb]{1, 0.8, 0.6} 62.7 & \cellcolor[rgb]{1, 0.8, 0.6} 61.3 & \cellcolor[rgb]{1, 0.9, 0.8} 62.6 & \cellcolor[rgb]{1, 0.8, 0.6} 66.1 & \cellcolor[rgb]{1, 0.8, 0.6} 64.2 & 63.7 & \cellcolor[rgb]{1, 0.8, 0.6} 68.8 & \cellcolor[rgb]{1, 0.8, 0.6} 65.4 \\
			\hline		
	\end{tabular}}
	\caption{Comparison of overall results on the ESC and CTB corpus. }
	\label{table:intra-cross}
	\vspace{-15pt}
\end{table*}

\subsection{Parameter Setting}
We use the pre-trained RoBERTa~\citep{Liu:et.al:arXiv:2019} model with 768-dimension base version provided by the HuggingFace transformers\footnote{\url{https://github.com/huggingface/transformers}} \citep{Wolf:et.al:EMNLP:2020}.
Our implementation is based on PyTorch framework\footnote{pytorch.org}, running on NVIDIA GTX 3090 GPUs.
The training process costs approximately 5 GPU hours on average.
We set the learning rate to 1e-5, batch size to 16.
The contrastive loss ratio $\beta$ is set to 0.5, the temperature parameter $\tau$ is set to 1.0, and the number of demonstrations is set to 4, viz. $(M, N) = (2, 2)$.
All trainable parameters are randomly initialized from normal distributions.

\subsection{Competitors}
We compare our ICCL with the following competitors: \textsf{ILP} \citep{Gao:et.al:NAACL:2019}, \textsf{KnowMMR} \citep{Liu:et.al:IJCAI:2021}, \textsf{RichGCN} \citep{Phu:et.al:NAACL:2021}, \textsf{CauSeRL} \citep{Zuo:et.al:arXiv:2021}, \textsf{LSIN} \citep{Cao:et.al:ACL:2021}, \textsf{LearnDA} \citep{Zuo-b:et.al:arXiv:2021}, \textsf{GESI} \citep{Fan:et.al:SIGIR:2022}, \textsf{ERGO} \citep{Chen:et.al:arXiv:2022}, \textsf{DPJL} \citep{Shen:et.al:COLING:2022}, \textsf{SemSln} \citep{Hu:et.al:arXiv:2023}.
The detailed introduction of competitors can be found in Appendix \ref{sec:comp}.

\section{Result and Analysis}
\subsection{Overall Result}
Table~\ref{table:intra-cross} compares the overall performance between our \textsf{ICCL} and the competitors on the ESC and CTB corpus.
We can  observe that the \textsf{ILP} cannot outperform other competitors, including the \textsf{RichGCN}, \textsf{GESI}, \textsf{ERGO}, and \textsf{SemSln}.
This can be attributed to their utilization of some graph neural networks for document structure encoding, enabling them to learn global contextual semantic for causality prediction.
We can also observe that the  \textsf{DPJL} adopting a kind of derivative prompt learning can significantly outperform the other competitors in intra-sentence causality identification. The outstanding performance can be attributed to its applying the prompt learning paradigm that transforms the ECI task to directly predict a PLM vocabulary word, other than fine-tuning a task-specific neural model upon a PLM. Although some other competitors have used external knowledge bases for relation identification, the prompt learning paradigm can better leverages potential causal knowledge in PLMs.

\par
Finally, our \textsf{ICCL} with different PLMs has achieved significant performance improvements overall competitors in terms of much higher F1 score with all intra-sentence, inter-sentence, and overall event causality identification on both ESC and CTB corpus.  We attribute its outstanding performance to applying contrastive learning on in-context demonstrations, by which our ICCL can better distinguish the semantic of causal and non-causal event pairs for causality prediction. Furthermore, we can also observe that using different PLMs do result in some performance variations, which are further discussed in Appendix~\ref{sec:plms}.
Finally the ICCL based on RoBERTa has achieved the best performance, as such we implement the remaining ablation experiments with RoBERTa.

\begin{table*}[ht]
	\centering
	\resizebox{0.9\textwidth}{!}{
		\renewcommand\arraystretch{1.1}
		
		\begin{tabular}{l|ccc|ccc|ccc|ccc}
			\hline
			\multirow{3}{*}{\textbf{Model}} & \multicolumn{9}{c|}{\textbf{EventStoryLine}} & \multicolumn{3}{c}{\textbf{Cause-TimeBank}} \\
			\cline{2-13}
			& \multicolumn{3}{c|}{Intra} & \multicolumn{3}{c|}{Cross} & \multicolumn{3}{c|}{Intra and Cross} & \multicolumn{3}{c}{Intra} \\
			\cline{2-13}
			& p (\%) & r (\%) & f1 (\%) & p (\%) & r (\%) & f1 (\%) & p (\%) & r (\%) & f1 (\%) & p (\%) & r (\%) & f1 (\%) \\
			\hline
			Prompt & 67.2 & 69.7 & 68.2 & \cellcolor[rgb]{1, 0.9, 0.8} 58.6 & 59.8 & 59.0 & \cellcolor[rgb]{1, 0.9, 0.8} 61.3 & 62.9 & 61.7 & 58.9 & 55.3 & 56.6 \\
			
			{In-context}& 66.0 & 72.4 & 68.9 & 57.7 & 60.9 & 59.1 & 60.4 & 64.5 & 62.2 & 60.3 & 58.0 & 58.7 \\
			
			ProCon w/o Demos & 60.8 & \cellcolor[rgb]{1, 0.9, 0.8} 77.9 & 68.2 & 54.2 & \cellcolor[rgb]{1, 0.9, 0.8} 65.6 & 59.3 & 56.4 & \cellcolor[rgb]{1, 0.9, 0.8} 69.4 & 62.1 & 51.5 & \cellcolor[rgb]{1, 0.8, 0.6} 71.8 & 58.9 \\
			
			ProCon w/ Demos & \cellcolor[rgb]{1, 0.9, 0.8} 67.1 & 73.5 & \cellcolor[rgb]{1, 0.9, 0.8} 70.1 & 58.0 & 61.9 & \cellcolor[rgb]{1, 0.9, 0.8} 59.8 & 60.9 & 64.5 & \cellcolor[rgb]{1, 0.9, 0.8} 63.1 & \cellcolor[rgb]{1, 0.8, 0.6} 66.9 & 60.2 & \cellcolor[rgb]{1, 0.9, 0.8} 62.5 \\
			
			EvtCon & 62.1 & \cellcolor[rgb]{1, 0.8, 0.6} 78.2 & 69.0 & 52.3 & \cellcolor[rgb]{1, 0.8, 0.6} 68.9 & 59.1 & 55.3 & \cellcolor[rgb]{1, 0.8, 0.6} 71.8 & 62.1 & 55.8 & 65.6 & 59.8 \\
			
			\hline
			\textbf{ICCL} & \cellcolor[rgb]{1, 0.8, 0.6} 67.5 & 73.7 & \cellcolor[rgb]{1, 0.8, 0.6} 70.4 & \cellcolor[rgb]{1, 0.8, 0.6} 60.3 & 62.7 & \cellcolor[rgb]{1, 0.8, 0.6} 61.3 & \cellcolor[rgb]{1, 0.8, 0.6} 62.6 & 66.1 & \cellcolor[rgb]{1, 0.8, 0.6} 64.2 & \cellcolor[rgb]{1, 0.9, 0.8} 63.7 & \cellcolor[rgb]{1, 0.9, 0.8} 68.8 & \cellcolor[rgb]{1, 0.8, 0.6} 65.4 \\
			\hline		
	\end{tabular}}
	\caption{Results of ablation study on the ESC and CTB corpus. }
	\label{table:ablation}
\end{table*}

\subsection{Ablation Study}
To examine the effectiveness of contrastive learning and in-context learning, we design the following ablation study. Table~\ref{table:ablation} compares their perfomance.
\par
$\bullet$ \textsf{Prompt} is prompt learning model, without demonstrations or contrastive module. 
\par
$\bullet$ \textsf{In-context} is in-context learning model, including retrieved demonstrations but without contrastive module. 
\par
$\bullet$ \textsf{ProCon w/o Demos} is prompt based contrastive model, but without demonstrations. We select positive and negative samples within batch insted of demonstrations, and use hidden state of $\mathtt{[MASK]}$ as input to contrastive module. 
\par
$\bullet$ \textsf{ProCon w/ Demos} is in-context based contrastive model with retrieved demonstrations, but still use the hidden state of $\mathtt{[MASK]}$ as input to contrastive module. 
\par
$\bullet$ \textsf{EvtCon} is event based prompt contrastive model, the only difference with \textsf{ProCon w/o Demos} is using hidden states of event pairs as contrastive module inputs.

\par
\textbf{\textsf{In-context learning:}} The first observation is that models incorporating in-context learning perform better. For example, the three models, \textsf{In-context}, \textsf{ProCon w/ Demos} and \textsf{ICCL} outperform \textsf{Prompt}, \textsf{ProCon w/o Demos} and \textsf{EvtCon}, respectively. This indicates the inclusion of demonstrations to explicitly guide the label prediction is highly effective in improving model performance. Furthermore, models with in-context learning show notable performance gains in challeging cross-sentence causality identification. That's because randomly selected demonstrations are predominantly composed of cross-sentence samples, which are more abundant in datasets. Therefore, PLMs develop a more comprehensive understanding of cross-sentence causality.

\par
\textbf{\textsf{Contrastive learning:}} We can observe that models with a contrastive module exhibit better performance. For example, both \textsf{ProCon w/ Demos} and \textsf{EvtCon} preform bette than \textsf{Prompt}. Additionally, both \textsf{ProCon w/o Demos} and \textsf{ICCL} preform bette than \textsf{In-context}. This can be attributed to the utilization of the contrastive learning paradigm, which enables the PLM to concentrate on event pairs or $\mathtt{[MASK]}$ and enhances PLM's ability to model them. Furthermore, it also helps discriminatively model positive and negative demonstrations, strengthening analogy between the query and all demonstrations. Additionally, we also observe that \textsf{EvtCon} usually outperformes \textsf{ProCon w/o Demos}. That's because hidden state of $\mathtt{[MASK]}$ serves as input for both contrastive and prediction module in the case of \textsf{ProCon w/o Demos}, yet the optimization directions of two modules do not completely align.

\subsection{Numbers of demonstrations}
\begin{figure}[h]
	\centering
	\subfloat[F1 score]{\begin{minipage}[t]{\linewidth}
			\centering
			\includegraphics[width=\textwidth , trim=14 0 32 10, clip]{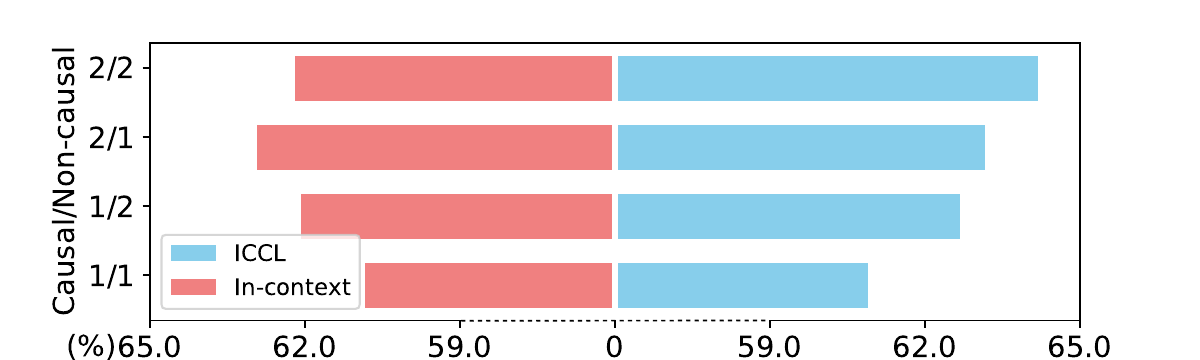}
	\end{minipage}} \\
	\vspace{-2pt}
	\subfloat[Recall]{\begin{minipage}[t]{\linewidth}
			\centering
			\includegraphics[width=\textwidth , trim=14 0 32 10, clip]{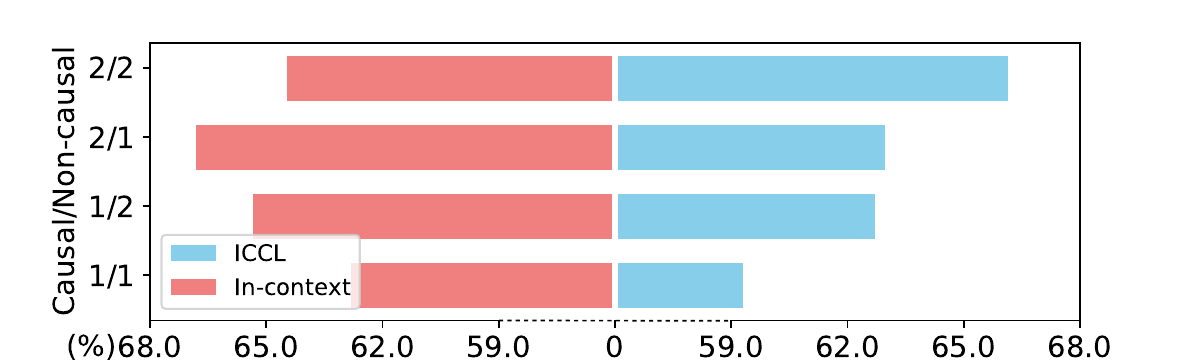}
	\end{minipage}} \\
	
	\vspace{-2pt}
	\subfloat[Precision]{\begin{minipage}[t]{\linewidth}
			\centering
			\includegraphics[width=\textwidth , trim=14 0 32 10, clip]{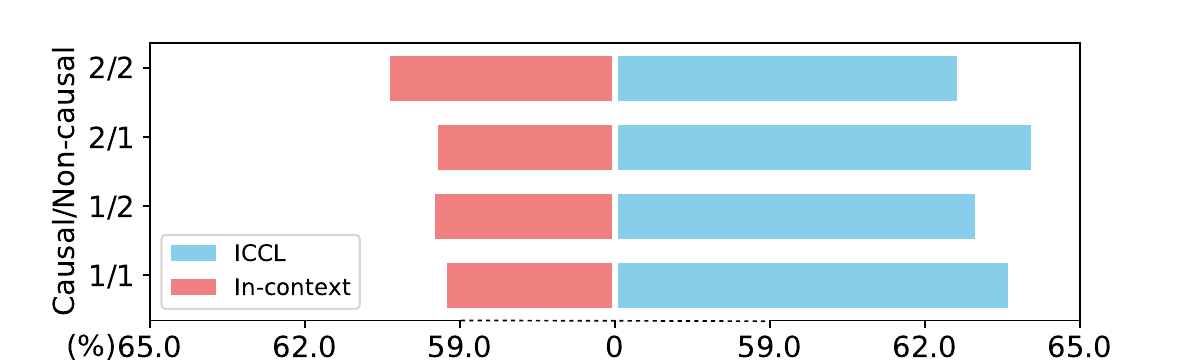}
	\end{minipage}}
	\vspace{-10pt}
	\caption{Comparision of \textsf{ICCL} and \textsf{In-context} model when using differenr numbers of causal and non-causal demonstrations on ESC corpus.}
	\label{fig:numbers}
\end{figure}

To further investigate the impact of demonstrations, we conducted an experiment that compared the performance of \textsf{In-context} and \textsf{ICCL} with varying numbers of causal and non-causal demonstrations. The results are showcased in Fig.~\ref{fig:numbers}.
\par

With more demonstrations, F1-score of both models initially exhibited improved performance, further validating the effectiveness of using demonstrations as explicit guidance. However, as the input length becomes too long, performance of \textsf{In-context} declines, while the performance of \textsf{ICCL} continues to improve. This can be attributed to the effectiveness of contrastive module used in \textsf{ICCL}, which aids the PLM in better focusing on event pairs, even with longer input. Additionally, the causal/non-causal ratio of 2/1 performs better compared to that of 1/2. That's because the dataset contains a limited number of causal samples. Increasing the number of causal demonstrations helps the model better learn the features of causal examples, mitigating the data imbalance issue.
\par

We can also observe that performance metrics of \textsf{{In-context}} model, particularly precision, exhibit minimal changes when the number of demonstrations varies. While as for our \textsf{ICCL} model, the precision and recall vary based on the ratio of causal and non-causal demonstrations. More non-causal demonstrations results in higher recall, while the opposite scenario leads to higher precision. These findings emphasize that the critical role of the contrastive module in enhancing analogy and enabling the PLM to effectively utilize positive and negative demonstrations.

\subsection{Few shot}
\begin{figure}[t]
	\centering
	\includegraphics[width=0.5\textwidth]{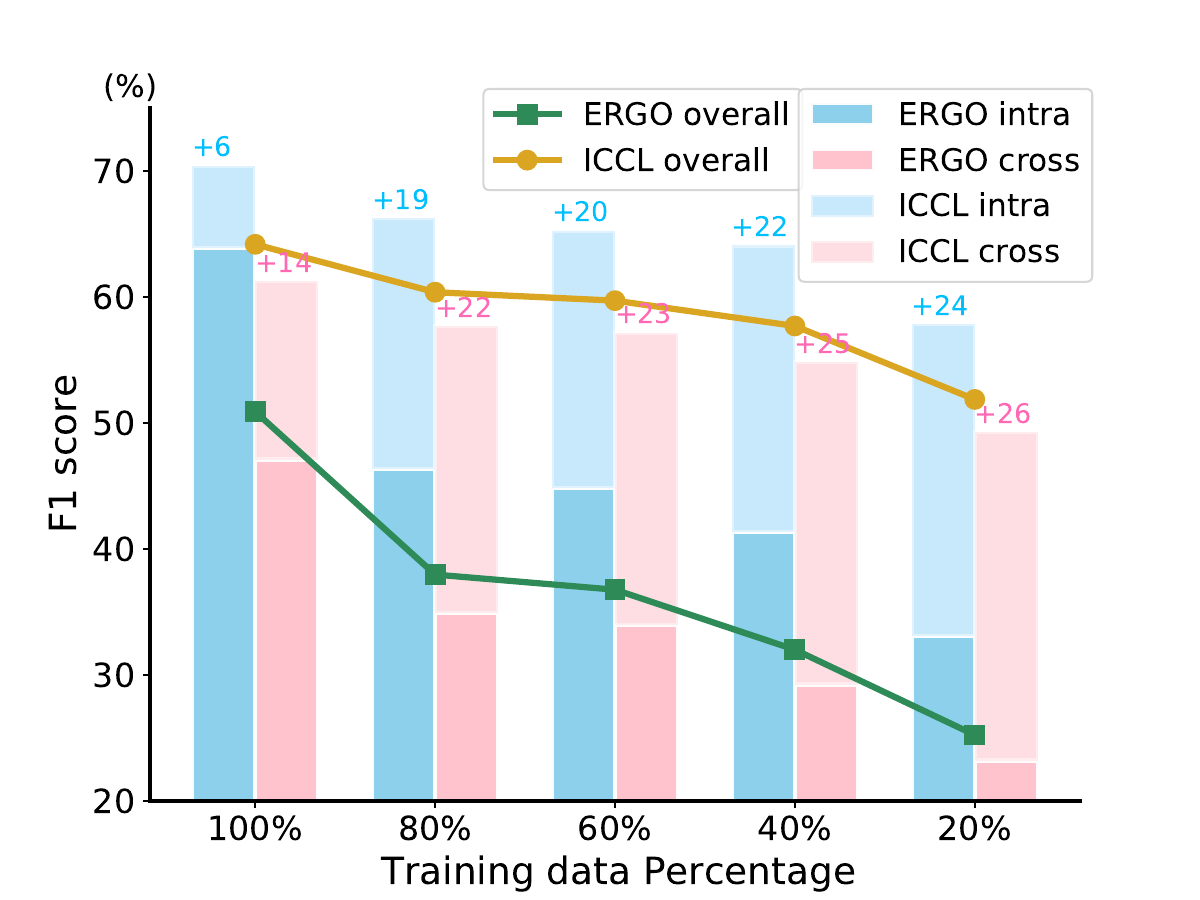}
	\caption{Results of few shot on ESC corpus. We replicated ERGO and get its few-shot results in the figure.}
	\label{fig:few}
\end{figure}

Some researchers have reported the robustness of prompt paradigm in using fewer training data \citep{Wang:et.al:EMNLP:2021, Ding:et.al:arXiv:2021}. Since our \textsf{ICCL} also employs a prompt-based method to predict the label, we examine its performance in low-resource scenarios and replicate the performance of \textsf{ERGO} as a benchmark for comparison. Fig.~\ref{fig:few} shows the performance comparison between \textsf{ERGO} and our \textsf{ICCL} on ESC corpus.
\par

As expected, the performance of \textsf{ICCL} gradually decreases as the amount of training data decreases. However, the decrease in performance is relatively slow, with an F1 score decrease of about 10\% when training data is reduced by 80\%, whereas the performance of \textsf{ERGO} declined by nearly 25\%. Notably, even with only 20\% of the training data, \textsf{ICCL} (F1: 51.9\%) outperformes \textsf{ERGO} (F1: 50.9\%) and many other competitors with full training data. These results confirm the effectiveness of \textsf{ICCL} even with fewer training data.

\begin{table}[ht]
	\centering
	\resizebox{\columnwidth}{!}{
		\renewcommand\arraystretch{1.2}
		
		\begin{tabular}{c|ccc|ccc}
			\hline
			\multirow{2}{*}{\textbf{Model}} & \multicolumn{3}{c|}{\textbf{EventStoryLine}} & \multicolumn{3}{c}{\textbf{Cause-TimeBank}} \\
			\cline{2-7}
			& P (\%) & R (\%) & F1 (\%) & P (\%)& R (\%)& F1 (\%)\\
			\hline
			BERT \citep{Gao:et.al:arXiv:2023} & 38.1 & 56.8 & 45.6 & 41.1 & 45.8 & 43.5\\
			RoBERTa \citep{Gao:et.al:arXiv:2023} & 42.1 & 64.0 & 50.8 & 39.9 & 60.9 & 48.2\\
			\hline
			T5 (Our implementation) & 36.2 & 49.2 & 40.7 & 7.7 & 52.1 & 12.1\\
			\hline
			gpt-3.5-turbo \citep{Gao:et.al:arXiv:2023} & 27.6 & 80.2 & 41.0 & 6.9 & 82.6 & 12.8\\
			gpt-4 \citep{Gao:et.al:arXiv:2023} & 27.2 & 94.7 & 42.2 & 6.1 &  97.4 & 11.5\\
			\hline
	\end{tabular}}
	\caption{Intra-sentence causality identification results of different PLMs and LLMs on the ESC and CTB corpus.}
	\label{table:intra}
\end{table}
\par

We also showcase the intra-sentence causality identification performance among different PLMs and several zero-shot models in the Table~\ref{table:intra}. We can not only find that our fine-tuned generative model, T5 (Our implementation), perform significantly worse than autoencoder models like BERT\textit{\small{-base}} \citep{Gao:et.al:arXiv:2023} and RoBERTa\textit{\small{-base}} \citep{Gao:et.al:arXiv:2023}, which confirms the conclusion drawn by \citet{Gao:et.al:arXiv:2023} that generative models may not be well-suited for causal reasoning tasks like ECI. We can also observe that although the ChatGPT models, such as \textsf{gpt-3.5-turbo} and \textsf{gpt-4},  have more comprehensive pre-training and larger model scales, these zero-shot models exhibit a significant performance gap compared to fine-tuned models like T5\textit{\small{-base}} and et al. This demonstrates the importance of fine-tune, indicating that it is challenging to address causal reasoning tasks like ECI in a zero-shot scenario. For more detailed analysis, please refer to Appendix~\ref{sec:plms}.

\subsection{Embedding Visualization}

\begin{figure}
	\centering
	\subfloat[Prompt]{\begin{minipage}[t]{0.48\linewidth}
			\centering
			\includegraphics[width=1\textwidth, trim=88 80 72 85, clip]{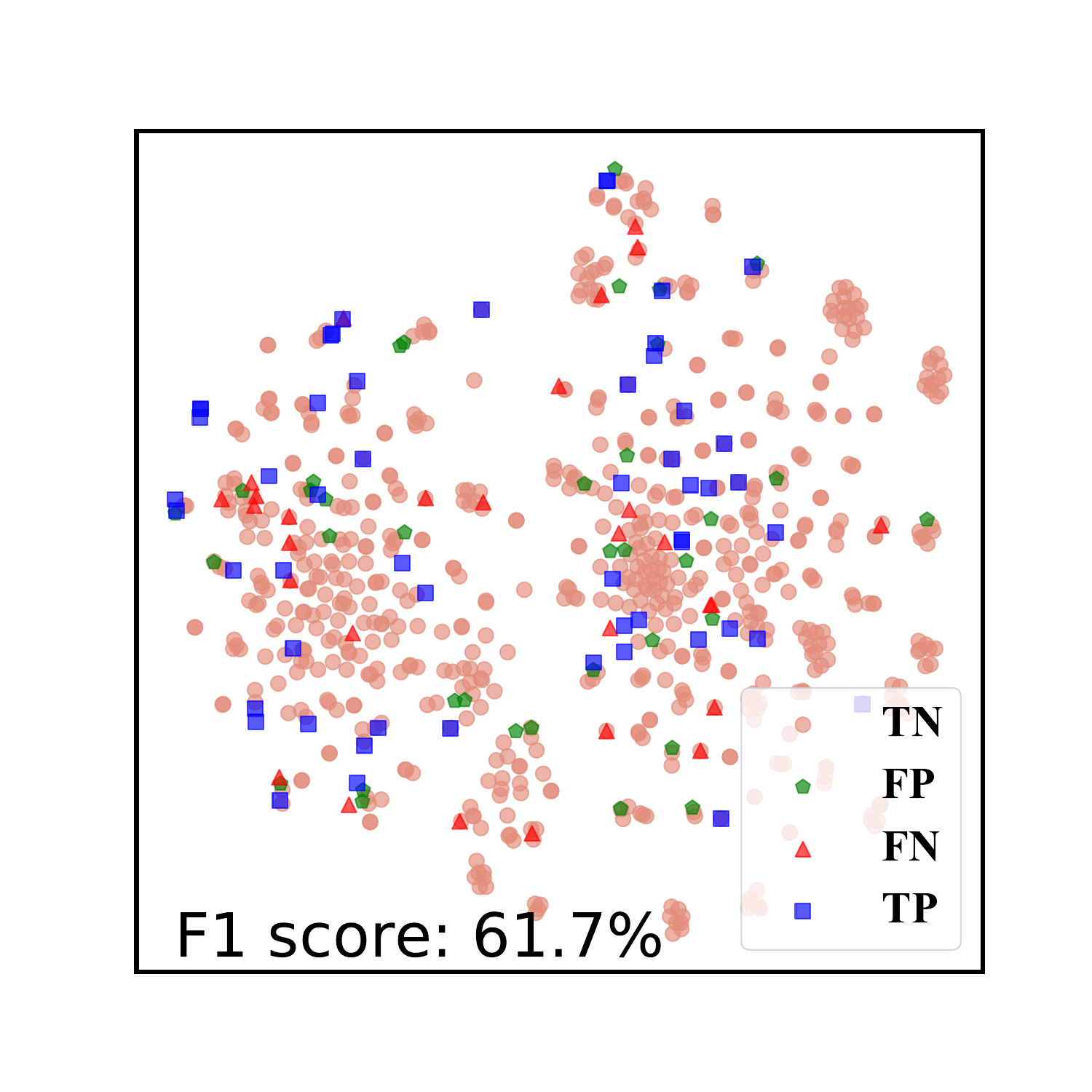}
	\end{minipage}}
	\hfill
	\subfloat[{In-context}]{\begin{minipage}[t]{0.48\linewidth}
			\centering
			\includegraphics[width=1\textwidth, trim=88 80 72 85, clip]{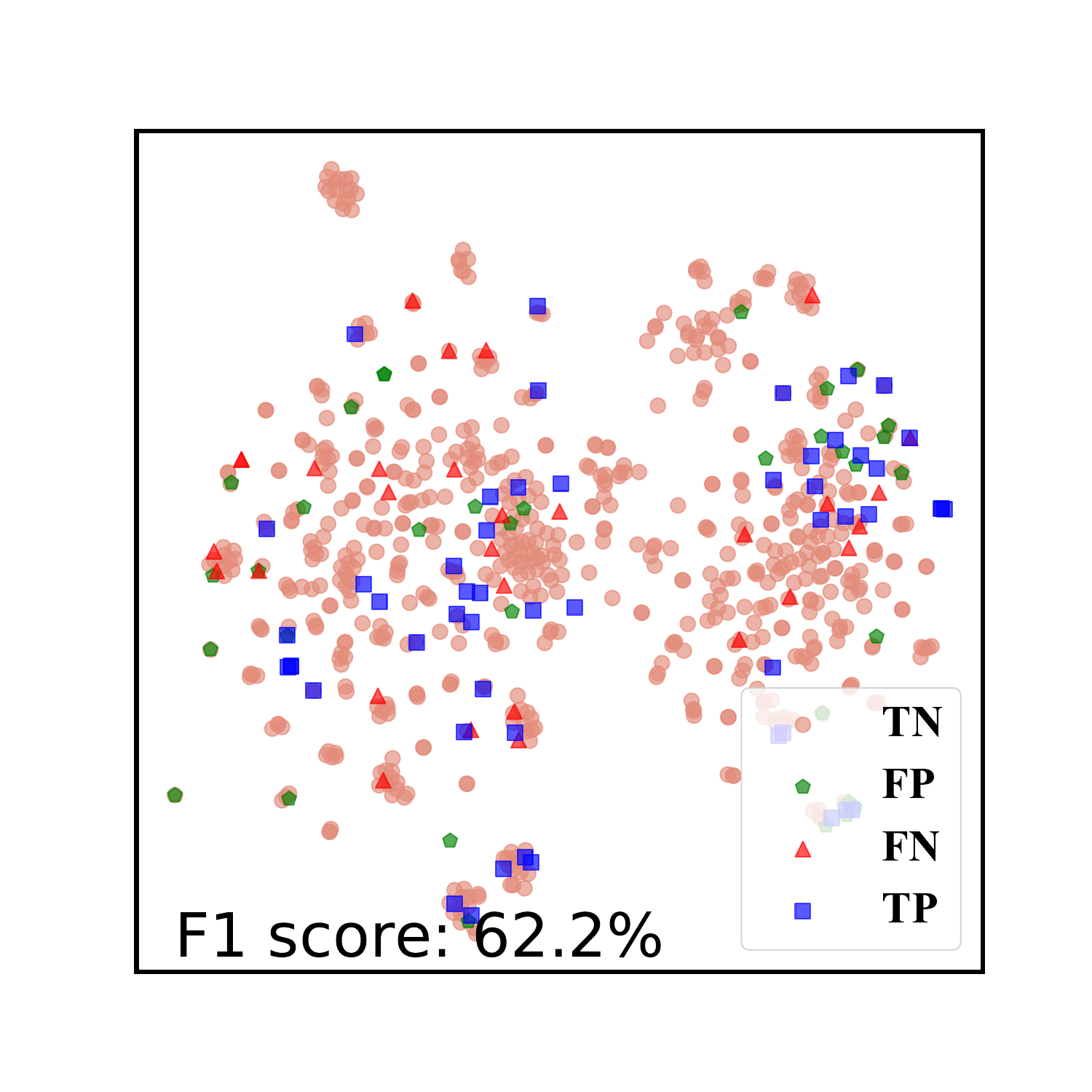}
	\end{minipage}} \\
	\subfloat[EvtCon]{\begin{minipage}[t]{0.48\linewidth}
			\centering
			\includegraphics[width=1\textwidth, trim=88 80 72 85, clip]{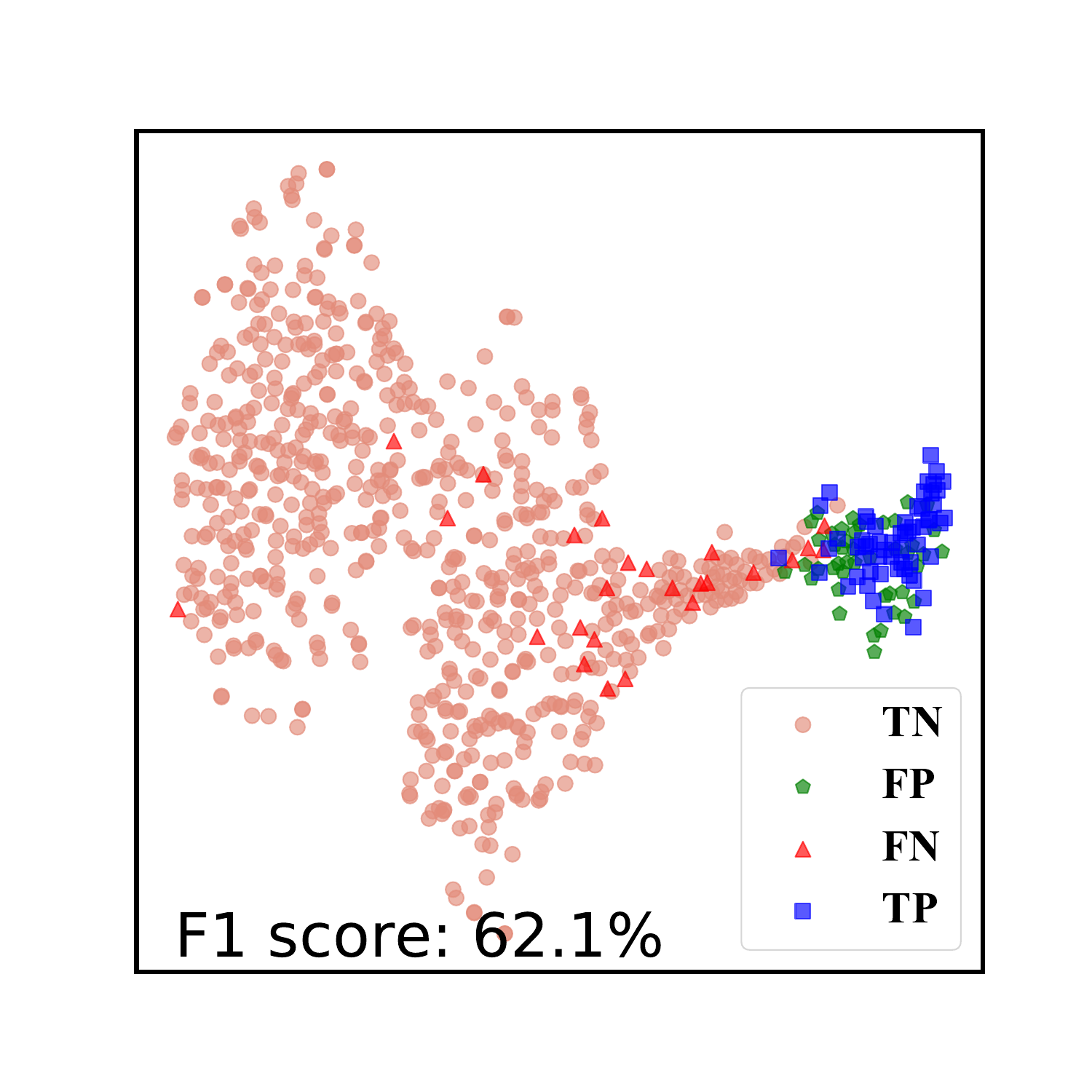}
	\end{minipage}}
	\hfill
	\subfloat[ICCL]{\begin{minipage}[t]{0.48\linewidth}
			\centering
			\includegraphics[width=1\textwidth, trim=88 80 72 85, clip]{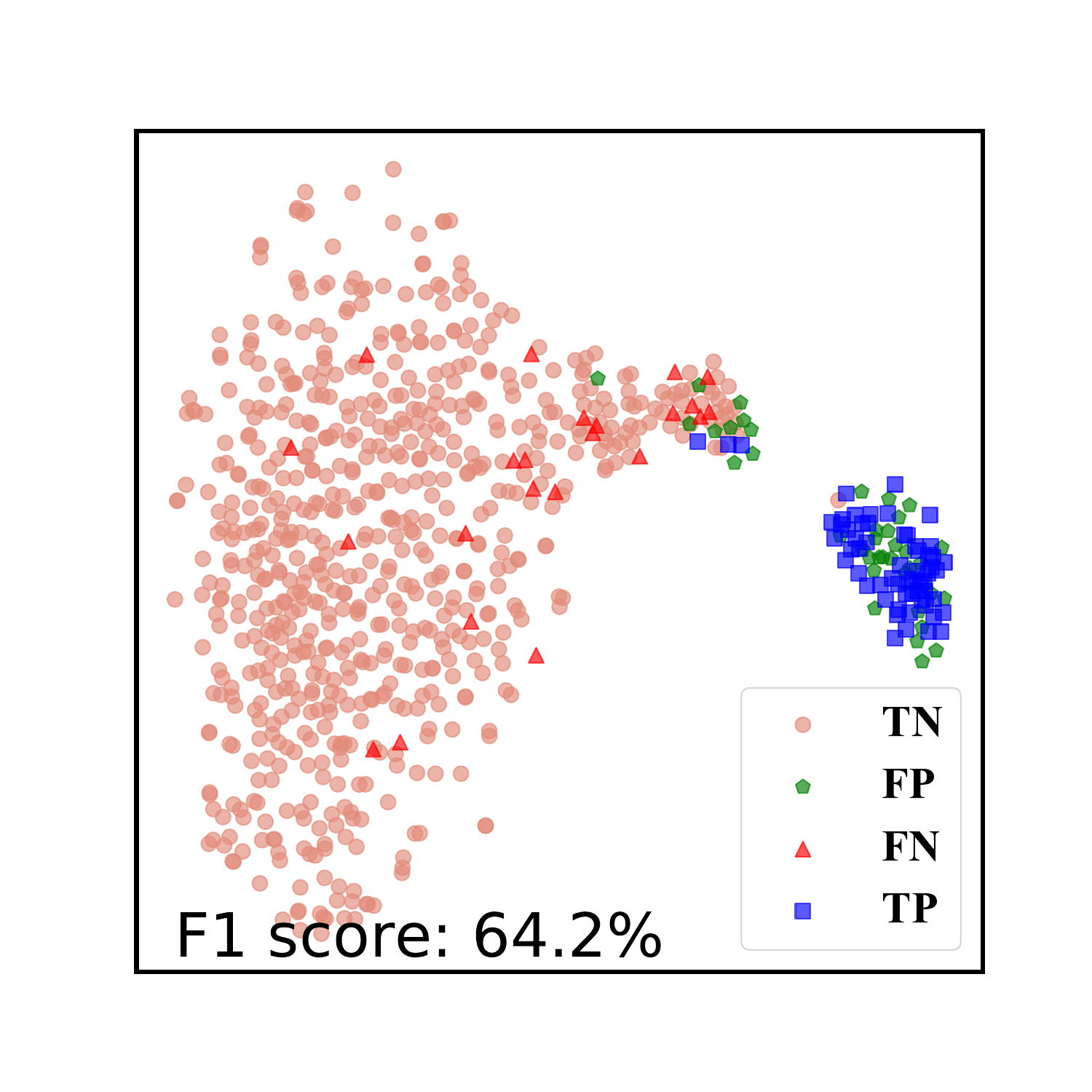}
	\end{minipage}}
	\caption{Visualization of the event pairs' embedding encoded by different models on ESC corpus}
	\label{fig:visualization}
\end{figure}

In order to verify the impact of contrastive module with event pairs as input, we compare the learned event pairs' embeddings $(h_{e_1}-h_{e_2})$ of different models on ESC test dataset by t-distributed stochastic neighbor embedding (t-SNE) \citep{Hinton:et.al:NeurIPS:2002}. In Fig.~\ref{fig:visualization}, we color-coded the points to represent \textit{True Nagetive (TN), False Positive (FP), False Nagetive (FN) and True Positive (TP)} samples.
\par

We can ovserve that models incorporating the contrastive module with event pairs as input exhibit a clear phenomenon of event pairs representations clustering together based on labels in the embedding space, which demonstrates the effective of the contrastive module. Additionally, representations of samples predicted to have the same label tended to cluster together, highlighting the crucial role of event pairs in identifying causality.

\section{Concluding Remarks}
In this paper, we propose an ICCL model and apply it on the ECI task. We leverage the causality knowledge of PLM by introducing explicit guidance through the inclusion of demonstrations, rather than relying on the design of complex prompts. Meanwhile, we employ contrastive learning with event pairs as input to enhance the PLM's attention to event pairs and strengthen the analogy between query and demonstrations. Experiments on the ESC and CTB corpus have validated that our ICCL can significantly outperform the state-of-the-art algorithms.
\par

In future, we will try to undertake experiments to apply our proposed framework to other NLP tasks in order to explore whether it can exhibit favorable adaptability when applied to different tasks.

\section*{Limitation}
Due to the input length limitations of the PLM, the number of demonstrations needs to be kept within a manageable range. However, our ICCL uses demonstrations as positive and negative samples in contrastive learning. This implies that there are limited positive and negative samples, which weakens the effectiveness of contrastive learning.

\section*{Acknowledgements}
This work is supported in part by National Natural Science Foundation of China (Grant No:
62172167). The computation is completed in the HPC Platform of Huazhong University of Science and Technology.

\section*{Ethics Statement}
This paper has no particular ethic consideration.

\bibliography{anthology,custom}

\begin{thebibliography}{40}
\expandafter\ifx\csname natexlab\endcsname\relax\def\natexlab#1{#1}\fi

\bibitem[{Balgi et~al.(2022)Balgi, Pena, and Daoud}]{Balgi:et.al:AAAI:2022}
Sourabh Balgi, Jose~M Pena, and Adel Daoud. 2022.
\newblock Personalized public policy analysis in social sciences using
  causal-graphical normalizing flows.
\newblock In \emph{Proceedings of the AAAI Conference on Artificial
  Intelligence}, volume~36, pages 11810--11818.

\bibitem[{Beltagy et~al.(2020)Beltagy, Peters, and
  Cohan}]{Beltagy:et.al:arXiv:2020}
Iz~Beltagy, Matthew~E Peters, and Arman Cohan. 2020.
\newblock Longformer: The long-document transformer.
\newblock \emph{arXiv preprint arXiv:2004.05150}.

\bibitem[{Berant et~al.(2014)Berant, Srikumar, Chen, Vander~Linden, Harding,
  Huang, Clark, and Manning}]{Berant:et.al:EMNLP:2014}
Jonathan Berant, Vivek Srikumar, Pei-Chun Chen, Abby Vander~Linden, Brittany
  Harding, Brad Huang, Peter Clark, and Christopher~D Manning. 2014.
\newblock Modeling biological processes for reading comprehension.
\newblock In \emph{Proceedings of the 2014 conference on empirical methods in
  natural language processing (EMNLP)}, pages 1499--1510.

\bibitem[{Breja and Jain(2020)}]{Breja:et.al:COLINS:2020}
Manvi Breja and Sanjay~Kumar Jain. 2020.
\newblock Causality for question answering.
\newblock In \emph{COLINS}, pages 884--893.

\bibitem[{Cao et~al.(2021)Cao, Zuo, Chen, Liu, Zhao, Chen, and
  Peng}]{Cao:et.al:ACL:2021}
Pengfei Cao, Xinyu Zuo, Yubo Chen, Kang Liu, Jun Zhao, Yuguang Chen, and Weihua
  Peng. 2021.
\newblock Knowledge-enriched event causality identification via latent
  structure induction networks.
\newblock In \emph{Proceedings of the 59th Annual Meeting of the Association
  for Computational Linguistics and the 11th International Joint Conference on
  Natural Language Processing (Volume 1: Long Papers)}, pages 4862--4872.

\bibitem[{Caselli and Vossen(2017)}]{Caselli:et.al:ACL:2017}
Tommaso Caselli and Piek Vossen. 2017.
\newblock The event storyline corpus: A new benchmark for causal and temporal
  relation extraction.
\newblock In \emph{Proceedings of the Events and Stories in the News Workshop},
  pages 77--86.

\bibitem[{Chen et~al.(2022)Chen, Cao, Deng, Li, Wang, Shao, and
  Zhang}]{Chen:et.al:arXiv:2022}
Meiqi Chen, Yixin Cao, Kunquan Deng, Mukai Li, Kun Wang, Jing Shao, and Yan
  Zhang. 2022.
\newblock Ergo: Event relational graph transformer for document-level event
  causality identification.
\newblock \emph{arXiv preprint arXiv:2204.07434}.

\bibitem[{Chen et~al.(2016)Chen, Zhu, Ling, Wei, Jiang, and
  Inkpen}]{Chen:et.al:arXiv:2016}
Qian Chen, Xiaodan Zhu, Zhenhua Ling, Si~Wei, Hui Jiang, and Diana Inkpen.
  2016.
\newblock Enhanced lstm for natural language inference.
\newblock \emph{arXiv preprint arXiv:1609.06038}.

\bibitem[{Devlin et~al.(2018)Devlin, Chang, Lee, and
  Toutanova}]{Devlin:et.al:arXiv:2018}
Jacob Devlin, Ming-Wei Chang, Kenton Lee, and Kristina Toutanova. 2018.
\newblock Bert: Pre-training of deep bidirectional transformers for language
  understanding.
\newblock \emph{arXiv preprint arXiv:1810.04805}.

\bibitem[{Ding et~al.(2021)Ding, Chen, Han, Xu, Xie, Zheng, Liu, Li, and
  Kim}]{Ding:et.al:arXiv:2021}
Ning Ding, Yulin Chen, Xu~Han, Guangwei Xu, Pengjun Xie, Hai-Tao Zheng, Zhiyuan
  Liu, Juanzi Li, and Hong-Gee Kim. 2021.
\newblock Prompt-learning for fine-grained entity typing.
\newblock \emph{arXiv preprint arXiv:2108.10604}.

\bibitem[{Dong et~al.(2022)Dong, Li, Dai, Zheng, Wu, Chang, Sun, Xu, and
  Sui}]{Dong:et.al:arXiv:2022}
Qingxiu Dong, Lei Li, Damai Dai, Ce~Zheng, Zhiyong Wu, Baobao Chang, Xu~Sun,
  Jingjing Xu, and Zhifang Sui. 2022.
\newblock A survey for in-context learning.
\newblock \emph{arXiv preprint arXiv:2301.00234}.

\bibitem[{Fan et~al.(2022)Fan, Liu, Qin, Zhang, and Xu}]{Fan:et.al:SIGIR:2022}
Chuang Fan, Daoxing Liu, Libo Qin, Yue Zhang, and Ruifeng Xu. 2022.
\newblock Towards event-level causal relation identification.
\newblock In \emph{Proceedings of the 45th International ACM SIGIR Conference
  on Research and Development in Information Retrieval}, pages 1828--1833.

\bibitem[{Gao et~al.(2023)Gao, Ding, Qin, and Liu}]{Gao:et.al:arXiv:2023}
Jinglong Gao, Xiao Ding, Bing Qin, and Ting Liu. 2023.
\newblock Is chatgpt a good causal reasoner? a comprehensive evaluation.
\newblock \emph{arXiv preprint arXiv:2305.07375}.

\bibitem[{Gao et~al.(2019)Gao, Choubey, and Huang}]{Gao:et.al:NAACL:2019}
Lei Gao, Prafulla~Kumar Choubey, and Ruihong Huang. 2019.
\newblock Modeling document-level causal structures for event causal relation
  identification.
\newblock In \emph{Proceedings of the 2019 Conference of the North American
  Chapter of the Association for Computational Linguistics: Human Language
  Technologies, Volume 1 (Long and Short Papers)}, pages 1808--1817.

\bibitem[{He et~al.(2020)He, Liu, Gao, and Chen}]{He:et.al:arXiv:2020}
Pengcheng He, Xiaodong Liu, Jianfeng Gao, and Weizhu Chen. 2020.
\newblock Deberta: Decoding-enhanced bert with disentangled attention.
\newblock \emph{arXiv preprint arXiv:2006.03654}.

\bibitem[{Hinton and Roweis(2002)}]{Hinton:et.al:NeurIPS:2002}
Geoffrey~E Hinton and Sam Roweis. 2002.
\newblock Stochastic neighbor embedding.
\newblock \emph{Advances in neural information processing systems}, 15.

\bibitem[{Hu et~al.(2023)Hu, Li, Jin, Bai, Guan, Guo, and
  Cheng}]{Hu:et.al:arXiv:2023}
Zhilei Hu, Zixuan Li, Xiaolong Jin, Long Bai, Saiping Guan, Jiafeng Guo, and
  Xueqi Cheng. 2023.
\newblock Semantic structure enhanced event causality identification.
\newblock \emph{arXiv preprint arXiv:2305.12792}.

\bibitem[{Khosla et~al.(2020)Khosla, Teterwak, Wang, Sarna, Tian, Isola,
  Maschinot, Liu, and Krishnan}]{Khosla:et.al:NeurIPS:2020}
Prannay Khosla, Piotr Teterwak, Chen Wang, Aaron Sarna, Yonglong Tian, Phillip
  Isola, Aaron Maschinot, Ce~Liu, and Dilip Krishnan. 2020.
\newblock Supervised contrastive learning.
\newblock \emph{Advances in neural information processing systems},
  33:18661--18673.

\bibitem[{Liu et~al.(2021{\natexlab{a}})Liu, Shen, Zhang, Dolan, Carin, and
  Chen}]{Liu:et.al:arXiv:2021}
Jiachang Liu, Dinghan Shen, Yizhe Zhang, Bill Dolan, Lawrence Carin, and Weizhu
  Chen. 2021{\natexlab{a}}.
\newblock What makes good in-context examples for gpt-$3 $?
\newblock \emph{arXiv preprint arXiv:2101.06804}.

\bibitem[{Liu et~al.(2021{\natexlab{b}})Liu, Chen, and
  Zhao}]{Liu:et.al:IJCAI:2021}
Jian Liu, Yubo Chen, and Jun Zhao. 2021{\natexlab{b}}.
\newblock Knowledge enhanced event causality identification with mention
  masking generalizations.
\newblock In \emph{Proceedings of the Twenty-Ninth International Conference on
  International Joint Conferences on Artificial Intelligence}, pages
  3608--3614.

\bibitem[{Liu et~al.(2023)Liu, Yuan, Fu, Jiang, Hayashi, and
  Neubig}]{Liu:ey.al:ACM:2023}
Pengfei Liu, Weizhe Yuan, Jinlan Fu, Zhengbao Jiang, Hiroaki Hayashi, and
  Graham Neubig. 2023.
\newblock Pre-train, prompt, and predict: A systematic survey of prompting
  methods in natural language processing.
\newblock \emph{ACM Computing Surveys}, 55(9):1--35.

\bibitem[{Liu et~al.(2019)Liu, Ott, Goyal, Du, Joshi, Chen, Levy, Lewis,
  Zettlemoyer, and Stoyanov}]{Liu:et.al:arXiv:2019}
Yinhan Liu, Myle Ott, Naman Goyal, Jingfei Du, Mandar Joshi, Danqi Chen, Omer
  Levy, Mike Lewis, Luke Zettlemoyer, and Veselin Stoyanov. 2019.
\newblock Roberta: A robustly optimized bert pretraining approach.
\newblock \emph{arXiv preprint arXiv:1907.11692}.

\bibitem[{Loshchilov and Hutter(2017)}]{Loshchilov:et.al:arXiv:2017}
Ilya Loshchilov and Frank Hutter. 2017.
\newblock Decoupled weight decay regularization.
\newblock \emph{arXiv preprint arXiv:1711.05101}.

\bibitem[{Mikolov et~al.(2013)Mikolov, Chen, Corrado, and
  Dean}]{Mikolov:et.al:arXiv:2013}
Tomas Mikolov, Kai Chen, Greg Corrado, and Jeffrey Dean. 2013.
\newblock Efficient estimation of word representations in vector space.
\newblock \emph{arXiv preprint arXiv:1301.3781}.

\bibitem[{Mirza and Tonelli(2014)}]{Mirza:et.al:COLING:2014}
Paramita Mirza and Sara Tonelli. 2014.
\newblock An analysis of causality between events and its relation to temporal
  information.
\newblock In \emph{Proceedings of COLING 2014, the 25th International
  Conference on Computational Linguistics: Technical Papers}, pages 2097--2106.

\bibitem[{Pennington et~al.(2014)Pennington, Socher, and
  Manning}]{Pennington:et.al:EMNLP:2014}
Jeffrey Pennington, Richard Socher, and Christopher~D Manning. 2014.
\newblock Glove: Global vectors for word representation.
\newblock In \emph{Proceedings of the 2014 conference on empirical methods in
  natural language processing (EMNLP)}, pages 1532--1543.

\bibitem[{Phu and Nguyen(2021)}]{Phu:et.al:NAACL:2021}
Minh~Tran Phu and Thien~Huu Nguyen. 2021.
\newblock Graph convolutional networks for event causality identification with
  rich document-level structures.
\newblock In \emph{Proceedings of the 2021 conference of the North American
  chapter of the association for computational linguistics: Human language
  technologies}, pages 3480--3490.

\bibitem[{Preethi et~al.(2015)Preethi, Uma et~al.}]{Preethi:et.al:science:2015}
Peter~G Preethi, Vilma Uma, et~al. 2015.
\newblock Temporal sentiment analysis and causal rules extraction from tweets
  for event prediction.
\newblock \emph{Procedia computer science}, 48:84--89.

\bibitem[{Pu et~al.(2023)Pu, Li, Wang, Li, Zheng, and Liao}]{Pu:et.al:ACL:2023}
Ruili Pu, Yang Li, Suge Wang, Deyu Li, Jianxing Zheng, and Jian Liao. 2023.
\newblock Enhancing event causality identification with event causal label and
  event pair interaction graph.
\newblock In \emph{Findings of the Association for Computational Linguistics:
  ACL 2023}, pages 10314--10322.

\bibitem[{Radinsky et~al.(2012)Radinsky, Davidovich, and
  Markovitch}]{Radinsky:et.al:WWW:2012}
Kira Radinsky, Sagie Davidovich, and Shaul Markovitch. 2012.
\newblock Learning causality for news events prediction.
\newblock In \emph{Proceedings of the 21st international conference on World
  Wide Web}, pages 909--918.

\bibitem[{Raffel et~al.(2020)Raffel, Shazeer, Roberts, Lee, Narang, Matena,
  Zhou, Li, and Liu}]{Raffel:et.al:JMLR:2020}
Colin Raffel, Noam Shazeer, Adam Roberts, Katherine Lee, Sharan Narang, Michael
  Matena, Yanqi Zhou, Wei Li, and Peter~J Liu. 2020.
\newblock Exploring the limits of transfer learning with a unified text-to-text
  transformer.
\newblock \emph{The Journal of Machine Learning Research}, 21(1):5485--5551.

\bibitem[{Shen et~al.(2022)Shen, Zhou, Wu, and Qi}]{Shen:et.al:COLING:2022}
Shirong Shen, Heng Zhou, Tongtong Wu, and Guilin Qi. 2022.
\newblock Event causality identification via derivative prompt joint learning.
\newblock In \emph{Proceedings of the 29th International Conference on
  Computational Linguistics}, pages 2288--2299.

\bibitem[{Speer et~al.(2017)Speer, Chin, and Havasi}]{Speer:et.al:AAAI:2017}
Robyn Speer, Joshua Chin, and Catherine Havasi. 2017.
\newblock Conceptnet 5.5: An open multilingual graph of general knowledge.
\newblock In \emph{Proceedings of the AAAI conference on artificial
  intelligence}, volume~31.

\bibitem[{Sun et~al.(2019)Sun, Wang, Li, Feng, Chen, Zhang, Tian, Zhu, Tian,
  and Wu}]{Sun:et.al:arXiv:2019}
Yu~Sun, Shuohuan Wang, Yukun Li, Shikun Feng, Xuyi Chen, Han Zhang, Xin Tian,
  Danxiang Zhu, Hao Tian, and Hua Wu. 2019.
\newblock Ernie: Enhanced representation through knowledge integration.
\newblock \emph{arXiv preprint arXiv:1904.09223}.

\bibitem[{Wang et~al.(2021)Wang, Wang, Qiu, Huang, and
  Gao}]{Wang:et.al:EMNLP:2021}
Chengyu Wang, Jianing Wang, Minghui Qiu, Jun Huang, and Ming Gao. 2021.
\newblock Transprompt: Towards an automatic transferable prompting framework
  for few-shot text classification.
\newblock In \emph{Proceedings of the 2021 conference on empirical methods in
  natural language processing}, pages 2792--2802.

\bibitem[{Wolf et~al.(2020)Wolf, Debut, Sanh, Chaumond, Delangue, Moi, Cistac,
  Rault, Louf, Funtowicz et~al.}]{Wolf:et.al:EMNLP:2020}
Thomas Wolf, Lysandre Debut, Victor Sanh, Julien Chaumond, Clement Delangue,
  Anthony Moi, Pierric Cistac, Tim Rault, R{\'e}mi Louf, Morgan Funtowicz,
  et~al. 2020.
\newblock Transformers: State-of-the-art natural language processing.
\newblock In \emph{Proceedings of the 2020 conference on empirical methods in
  natural language processing: system demonstrations}, pages 38--45.

\bibitem[{Xiang et~al.(2022)Xiang, Wang, Dai, and
  Wang}]{Xiang:et.al:COLING:2022}
Wei Xiang, Zhenglin Wang, Lu~Dai, and Bang Wang. 2022.
\newblock Connprompt: Connective-cloze prompt learning for implicit discourse
  relation recognition.
\newblock In \emph{Proceedings of the 29th International Conference on
  Computational Linguistics}, pages 902--911.

\bibitem[{Zhao et~al.(2021)Zhao, Ji, He, Liu, and
  Ren}]{Zhao:et.al:Sciences:2021}
Kun Zhao, Donghong Ji, Fazhi He, Yijiang Liu, and Yafeng Ren. 2021.
\newblock \href {https://doi.org/https://doi.org/10.1016/j.ins.2021.01.078}
  {Document-level event causality identification via graph inference
  mechanism}.
\newblock \emph{Information Sciences}, 561:115--129.

\bibitem[{Zuo et~al.(2021{\natexlab{a}})Zuo, Cao, Chen, Liu, Zhao, Peng, and
  Chen}]{Zuo:et.al:arXiv:2021}
Xinyu Zuo, Pengfei Cao, Yubo Chen, Kang Liu, Jun Zhao, Weihua Peng, and Yuguang
  Chen. 2021{\natexlab{a}}.
\newblock Improving event causality identification via self-supervised
  representation learning on external causal statement.
\newblock \emph{arXiv preprint arXiv:2106.01654}.

\bibitem[{Zuo et~al.(2021{\natexlab{b}})Zuo, Cao, Chen, Liu, Zhao, Peng, and
  Chen}]{Zuo-b:et.al:arXiv:2021}
Xinyu Zuo, Pengfei Cao, Yubo Chen, Kang Liu, Jun Zhao, Weihua Peng, and Yuguang
  Chen. 2021{\natexlab{b}}.
\newblock Learnda: Learnable knowledge-guided data augmentation for event
  causality identification.
\newblock \emph{arXiv preprint arXiv:2106.01649}.

\end{thebibliography}

\clearpage

\begin{table*}[t]
	\centering
	\resizebox{2\columnwidth}{!}{
		\renewcommand\arraystretch{1.1}
		
		\begin{tabular}{l|ccc|ccc|ccc|ccc}
			\hline
			\multirow{3}{*}{\textbf{Model}} & \multicolumn{9}{c|}{\textbf{EventStoryLine}} & \multicolumn{3}{c}{\textbf{Cause-TimeBank}} \\
			\cline{2-13}
			& \multicolumn{3}{c|}{Intra} & \multicolumn{3}{c|}{Cross} & \multicolumn{3}{c|}{Intra and Cross} & \multicolumn{3}{c}{Intra} \\
			\cline{2-13}
			& p (\%) & r (\%) & f1 (\%) & p (\%) & r (\%) & f1 (\%) & p (\%) & r (\%) & f1 (\%) & p (\%) & r (\%) & f1 (\%) \\
			\hline
			T5 & 36.2 & 49.2 & 40.7 & - & - & - & - & - & - & 7.7 & 52.1 & 12.1 \\
			BERT \textbf{\dag} & 38.1 & 56.8 & 45.6 & - & - & - & - & - & - & 41.4 & 45.8 & 43.5 \\
			RoBERTa \textbf{\dag} & 42.1 & 64.0 & 50.8 & - & - & - & - & - & - & 39.9 & 60.9 & 48.2 \\
			\hline
			text-davinci-002 \textbf{\dag} & 23.2 & 80.0 & 36.0 & - & - & - & - & - & - & 5.0 & 75.2 & 9.3 \\
			text-davinci-003 \textbf{\dag} & 33.2 & 74.4 & 45.9 & - & - & - & - & - & - & 8.5 & 64.4 & 15.0 \\
			gpt-3.5-turbo \textbf{\dag} & 27.6 & \cellcolor[rgb]{1, 0.9, 0.8} 80.2 & 41.0 & - & - & - & - & - & - & 6.9 & \cellcolor[rgb]{1, 0.9, 0.8} 82.6 & 12.8 \\
			gpt-4 \textbf{\dag} & 27.2 & \cellcolor[rgb]{1, 0.8, 0.6} 94.7 & 42.2 & - & - & - & - & - & - & 6.1 & \cellcolor[rgb]{1, 0.8, 0.6} 97.4 & 11.5 \\
			\hline
			In-context\textit{\small{-T5}} & 63.3 & 62.6 & 62.7 & 53.7 & 46.6 & 49.3 & 57.0 & 51.5 & 53.7 & 9.2 & 50.4 & 14.8 \\
			In-context\textit{\small{-RoBERTa}}& 66.0 & 72.4 & \cellcolor[rgb]{1, 0.9, 0.8} 68.9 & 57.7 & \cellcolor[rgb]{1, 0.9, 0.8} 60.9 & 59.1 & 60.4 & \cellcolor[rgb]{1, 0.9, 0.8} 64.5 & 62.2 & 60.3 & 58.0 & 58.7 \\
			\hline
			ILP \citep{Gao:et.al:NAACL:2019} & 38.8 & 52.4 & 44.6 & 35.1 & 48.2 & 40.6 & 36.2 & 49.5 & 41.9 & - & - & - \\
			KnowMMR \citep{Liu:et.al:IJCAI:2021} & 41.9 & 62.5 & 50.1 & - & - & - & - & - & - & 36.6 & 55.6 & 44.1 \\
			RichGCN \citep{Phu:et.al:NAACL:2021} & 49.2 & 63.0 & 55.2 & 39.2 & 45.7 & 42.2 & 42.6 & 51.3 & 46.6 & 39.7 & 56.5 & 46.7 \\
			CauSeRL \citep{Zuo:et.al:arXiv:2021} & 41.9 & 69.0 & 52.1 & - & - & - & - & - & - & 43.6 & 68.1 & 53.2 \\
			LSIN \citep{Cao:et.al:ACL:2021} & 47.9 & 58.1 & 52.5 & - & - & - & - & - & - & 51.5 & 56.2 & 53.7 \\
			LearnDA \citep{Zuo-b:et.al:arXiv:2021} & 42.2 & 69.8 & 52.6 & - & - & - & - & - & - & 41.9 & 68.0 & 51.9 \\
			GESI \citep{Fan:et.al:SIGIR:2022} & - & - & 50.3 & - & - & 49.3 & - & - & 49.4 & - & - & - \\
			ERGO \citep{Chen:et.al:arXiv:2022} & 57.5 & 72.0 & 63.9 & 51.6 & 43.3 & 47.1 & 48.6 & 53.4 & 50.9 & 62.1 & 61.3 & 61.7 \\
			DPJL \citep{Shen:et.al:COLING:2022} & 65.3 & 70.8 & 67.9 & - & - & - & - & - & - & 63.6 & 66.7 & 64.6 \\
			SemSln \citep{Hu:et.al:arXiv:2023} & 64.2 & 65.7 & 64.9 & - & - & - & - & - & - & 52.3 & 65.8 & 58.3 \\
			\hline
			ICCL\textit{\small{-BERT}} & 64.9 & 69.6 & 67.1 & 56.3 & 58.4 & 57.2 & 59.0 & 61.9 & 60.4 & 60.5 & 58.4 & 59.1 \\
			ICCL\textit{\small{-ERNIE}} & 66.8 & 68.5 & 67.5 & \cellcolor[rgb]{1, 0.8, 0.6} 63.7 & 56.2 & 59.5 & \cellcolor[rgb]{1, 0.8, 0.6} 64.8 & 60.0 & 62.1 & \cellcolor[rgb]{1, 0.9, 0.8} 64.8 & 66.0 & 64.7 \\
			ICCL\textit{\small{-DeBERTa}} & \cellcolor[rgb]{1, 0.8, 0.6} 67.6 & 73.7 & \cellcolor[rgb]{1, 0.8, 0.6}70.4 & \cellcolor[rgb]{1, 0.9, 0.8} 61.8 & 58.4 & \cellcolor[rgb]{1, 0.9, 0.8} 59.9 & 61.7 & 63.2 & \cellcolor[rgb]{1, 0.9, 0.8} 63.3 & \cellcolor[rgb]{1, 0.8, 0.6} 66.7 & 64.4 & \cellcolor[rgb]{1, 0.9, 0.8} 64.9 \\
			ICCL\textit{\small{-RoBERTa}} & \cellcolor[rgb]{1, 0.9, 0.8} 67.5 & 73.7 & \cellcolor[rgb]{1, 0.8, 0.6} 70.4 & 60.3 & \cellcolor[rgb]{1, 0.8, 0.6} 62.7 & \cellcolor[rgb]{1, 0.8, 0.6} 61.3 & \cellcolor[rgb]{1, 0.9, 0.8} 62.6 & \cellcolor[rgb]{1, 0.8, 0.6} 66.1 & \cellcolor[rgb]{1, 0.8, 0.6} 64.2 & 63.7 & 68.8 & \cellcolor[rgb]{1, 0.8, 0.6} 65.4 \\
			\hline		
	\end{tabular}}
	\caption{Comparison of overall results on the ESC and CTB corpus. Performance of models marked with "\textbf{\dag}" after the name are cited from the research of \citet{Gao:et.al:arXiv:2023}. We name our models in the format of {Model}\textit{\small{-PLM}}, for example, ICCL\textit{\small{-BERT}} is the version of ICCL model based on BERT.	}
	\label{table:PLMs}
\end{table*}

\appendix

\section{Study of PLMs}
\label{sec:plms}

The \textsf{ICCL} model we proposed is a PLM-sensitive model. In order to investigate the performance of our model using different PLMs and select the most suitable one, we conducted PLM ablation experiment to test performance of our model with differenr PLMs. Furthermore, we also cited performance of some baseline methods based on PLMs finetuned on full training datasets from the work of \citet{Gao:et.al:arXiv:2023} to evaluate various PLMs and summarized the results in Table~\ref{table:PLMs}. The introductions of main PLMs we considered are as follows:

\vspace{10pt}
$\bullet$ \textbf{BERT} \cite{Devlin:et.al:arXiv:2018}: The most representive PLM proposed by Google\footnote{\url{https://github.com/google-research/bert}}, which is pre-trained using a cloze task and a next
sentence prediction task.
\par
$\bullet$ \textbf{RoBERTa} \cite{Liu:et.al:arXiv:2019}: A BERT enhanced PLM proposed by Facebook\footnote{\url{https://github.com/pytorch/fairseq/}}, which removes the next sentence prediction objective and is pre-trained on a much larger dataset with some modified key hyper-parameters.
\par
$\bullet$ \textbf{ERNIE} \cite{Sun:et.al:arXiv:2019}: A knowledge enhaced PLM proposed by Baidu\footnote{\url{https://github.com/PaddlePaddle/ERNIE}}, which uses some knowledgeable masking strategies in pre-training.
\par
$\bullet$ \textbf{DeBERTa} \cite{He:et.al:arXiv:2020}: The latest masked PLM proposed by Microsoft\footnote{\url{https://github.com/microsoft/DeBERTa}}, which improves BERT and RoBERTa models using a disentangled attention mechanism and an enhanced mask decoder.
\par
$\bullet$ \textbf{T5} \cite{Raffel:et.al:JMLR:2020}: A generative language model proposed by Google\footnote{\url{https://github.com/google-research/multilingual-t5}} in 2020, which is pre-trained on large-scale unsupervised datasets using an autoregressive approach and fine-tuned on task-specific annotated data. It has achieved state-of-the-art performance on multiple NLP tasks such as text generation, summarization, and translation.
\par

As shown in Table~\ref{table:PLMs}, according to the research by \citet{Gao:et.al:arXiv:2023}, it can be observed that,our fine-tuned generative model, T5\textit{\small{-base}}, performs significantly worse than autoencoder models like BERT\textit{\small{-base}} \citep{Gao:et.al:arXiv:2023} and RoBERTa\textit{\small{-base}} \citep{Gao:et.al:arXiv:2023}. Moreover, the performance of \textsf{In-context\textit{\small{-T5}}} is also far inferior to the model \textsf{In-context\textit{\small{-RoBERTa}}}. This confirms the conclusion drawn by \citet{Gao:et.al:arXiv:2023} that generative models may not be well-suited for causal reasoning tasks like ECI. Additionally, although the ChatGPT models, such as \textsf{gpt-3.5-turbo}) and \textsf{gpt-4}, have more comprehensive pre-training and larger model scales, these zero-shot models exhibit a significant performance gap compared to fine-tuned models like T5\textit{\small{-base}} and et al. This demonstrates the importance of fine-tune, indicating that it is challenging to address causal reasoning tasks like ECI in a zero-shot scenario.
\par

Besides, we can observe that our \textsf{ICCL} with all four PLMs has achieved better performance than most of competitors on both ESC and CTB corpus. Even our \textsf{ICCL\textit{\small{-BERT}}} outperformed many competitors with advanced PLMs, such as \textsf{ERGO} based on Longformer\cite{Beltagy:et.al:arXiv:2020}. This further demonstrates the effectiveness of our proposed method. Compared to approaches involving complex prompts or joint training across multiple tasks, our approach of utilizing simple explicit guidance and leveraging it for contextual contrastive learning better harnesses the semantic knowledge embedded in PLMs and guides their understanding of causal relationships.
\par

We can also observe that using different PLMs do result in some performance variations. This is not unexpected. It can be attributed to that while all the four PLMs employ a kind of Transformer-based model in pre-training on large-scale corpus, their training strategies or training corpus are not entirely identical. Compared to \textsf{ICCL\textit{\small{-BERT}}}, our \textsf{ICCL} model using ERNIE, DeBERTa, or RoBERTa achieved better performance. This is attributed to the fact that these three PLMs have made some optimizations based on BERT. For example, ERNIE utilizes a strategy of continuous learning in the pre-training stage. Finally, \textsf{ICCL\textit{\small{-RoBERTa}}} achieved the best performance, which removes the next sentence prediction objective and is pre-trained on a much larger dataset with some modified key hyper-parameters. Therefore, we implement the remaining ablation experiments with RoBERTa.

\begin{figure*}[t]
	\centering
	\includegraphics[width=0.9\textwidth , trim=20 260 60 35, clip]{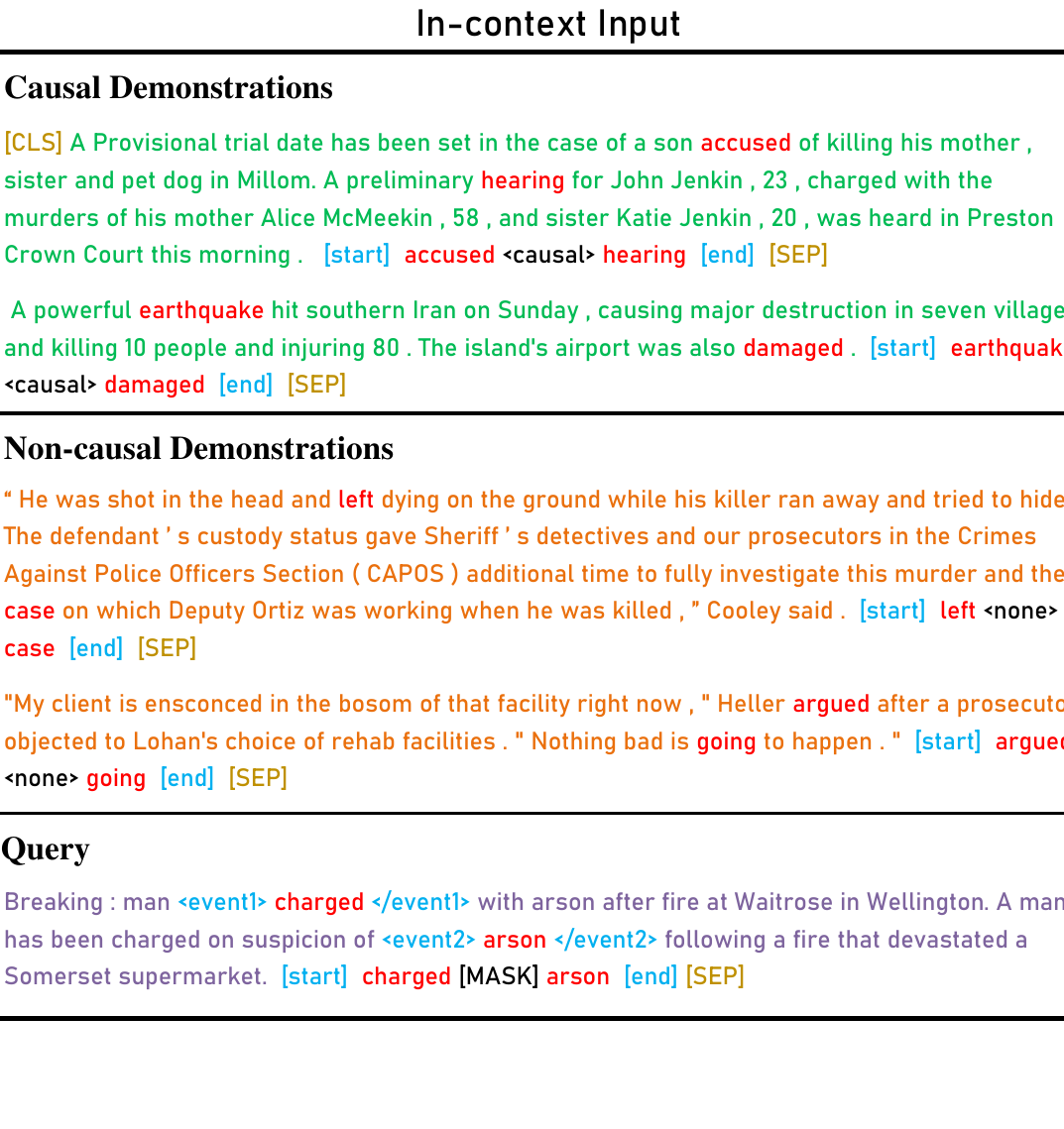}
	\vspace{-30pt}
	\caption{Example of in-context input. The line breaks and the title of each part (ex. Causal Demonstrations) are only to make the input readable, and they are not included in the actual input.}
	\label{fig:input}
\end{figure*}

\begin{figure}[t]
	\centering
	\includegraphics[width=0.5\textwidth , trim=6 12 0 0, clip]{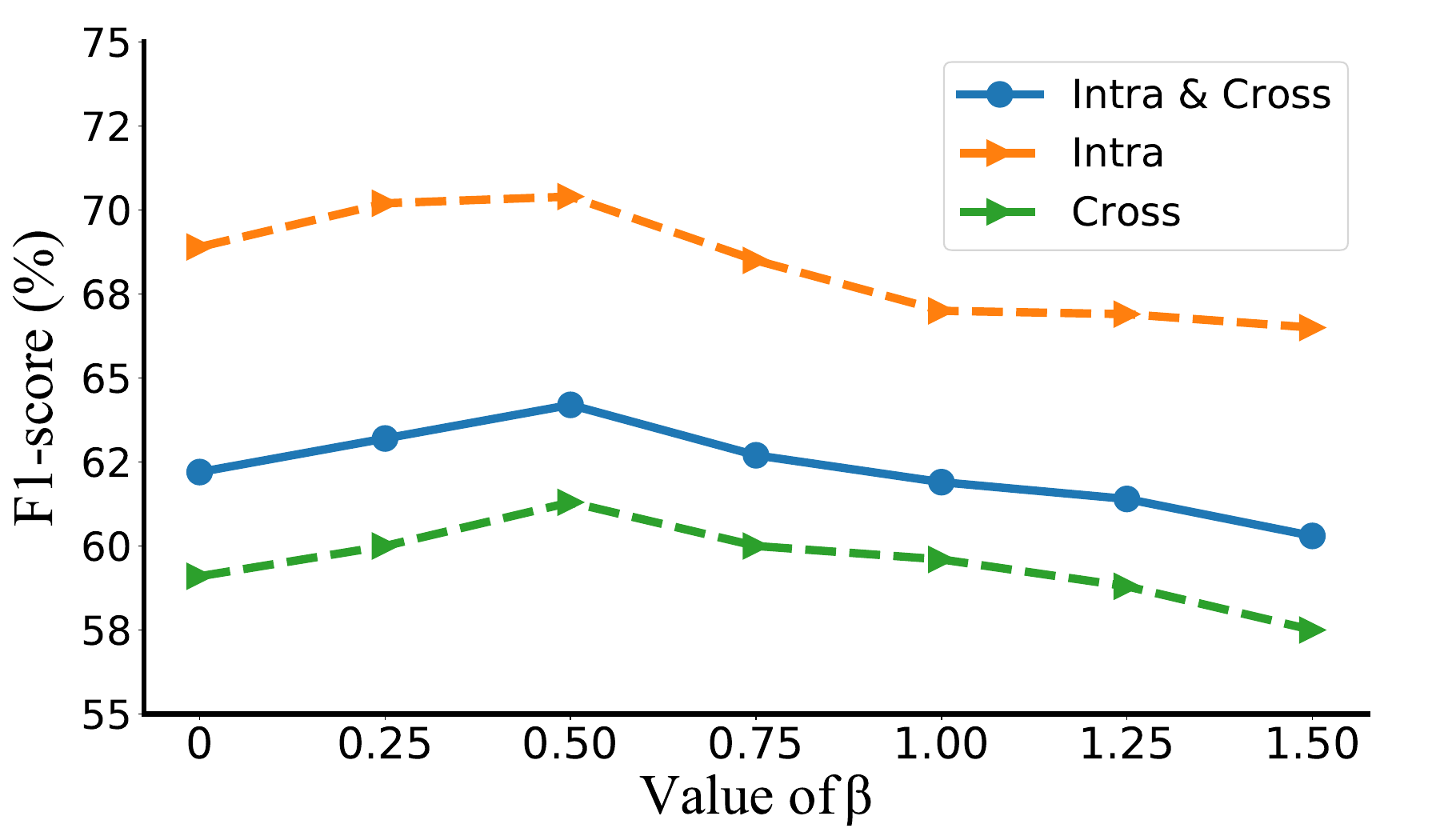}
	\caption{Comparision of ICCL model with different value of $\beta$ on the ESC corpus.}
	\label{fig:beta}
	\vspace{-10pt}
\end{figure}

\section{Competitors}
\label{sec:comp}
Table~\ref{table:PLMs} also presents results of more competitors. The introductions of these competitors are as follows:

\vspace{10pt}
$\bullet$ \textsf{ILP} \citep{Gao:et.al:NAACL:2019} employs integer linear programming to detect causal relationships by incorporating causal constraints at document level.
\par
$\bullet$ \textsf{KnowMMR} \citep{Liu:et.al:IJCAI:2021} utilizes external knowledge to extract event causality patterns.
\par
$\bullet$ \textsf{RichGCN} \citep{Phu:et.al:NAACL:2021} uses a graph convolutional network to learn context-enriched representations for event pairs based on document-level information.
\par
$\bullet$ \textsf{CauSeRL} \citep{Zuo:et.al:arXiv:2021} employs a contrastive approach to transfer externally learned causal statements.
\par
$\bullet$ \textsf{LSIN} \citep{Cao:et.al:ACL:2021} employs graph induction to acquire external structural and relational knowledge.
\par
$\bullet$ \textsf{LearnDA} \citep{Zuo-b:et.al:arXiv:2021} utilizes knowledge bases to interactively generate training data.
\par
$\bullet$ \textsf{GESI} \citep{Fan:et.al:SIGIR:2022} designs a graph convolutional network on an event co-reference graph to model causality.
\par
$\bullet$ \textsf{ERGO} \citep{Chen:et.al:arXiv:2022} constructs a relational graph where event pairs serve as nodes, capturing causal transitivity through a transformer-like network.
\par
$\bullet$ \textsf{DPJL} \citep{Shen:et.al:COLING:2022} leverages two derivative prompt tasks to identify causality.
\par
$\bullet$ \textsf{SemSln} \citep{Hu:et.al:arXiv:2023} uses a Graph Neural Network (GNN) to learn from event-centric structures for encoding events.
\vspace{10pt}

\section{In-context input}
\label{sec:input}
To help readers gain a better understanding of the in-context input generated by our Prompt module, we provide a specific example in Fig.~\ref{fig:input}.
\par
As depicted in Fig.~\ref{fig:input}, we randomly chose two causal demonstrations and two non-causal demonstrations from the training dataset for the query. Each segment in Fig.~\ref{fig:input} represents either a prompted demonstration or a prompted query. The initial two segments, highlighted in green font, represents demonstrations labeled as ${<causal>}$. The following two segments, highlighted in orange font, represents demonstrations labeled as ${<none>}$. Lastly, the final segment, highlighted in purple font, represents the query to predict.
\par

Besides, we have annotated some specific tokens we used with special colors. We utilized three PLM-special tokens: ${[CLS]}$ to indicate the beginning of the input, ${[SEP]}$ as a sentence separator, and ${[MASK]}$ as a placeholder for the label to predict. Furthermore, we have also devised some additional special tokens: ${[start]}$ and ${[end]}$ are used to indicate the beginning and end of the cloze template respectively, {${[event1]}$, ${[event1/]}$, ${[event2]}$, ${[event2/]}$} are used to highlight the events in the query, while ${<causal>}$ and ${<none>}$ respectivaly represent the causal and uncausal labels for the demonstrations.
\par

Additionally, although the contrastive module only works during the training phase, we select appropriate demonstrations for the query in both training and testing phases. Specifically, we randomly select $M$ samples labeled as ${<causal>}$ and $N$ samples labeled as ${<none>}$ from training dataset to be demonstrations. And on the contrastive learning process, positive demonstrations are those with the same label as the query, while negative demonstrations have different labels. Furthermore, during training phase, different demonstrations are retrieved for the same query in different epochs to introduce variability and enhance the model's ability to handle diverse instances of the same query. However, during validation and testing state, demonstrations retrieved for the same query, as well as the permutation order, remain consistent across epochs which ensures fair evaluation.

\section{Study of $\beta$}
\label{sec:beta}
To further explore how to balance the importance of contrastive loss and prediction loss, we investigated the performance of the ICCL model with different values of the hyperparameter $\beta$ on the ESC corpus.
\par
As shown in Fig.~\ref{fig:beta}, we can observe that as $\beta$ increases from 0, the performance of the model initially improves and then starts to decline. The optimal performance on both intra-sentence causality and cross-sentence causality is achieved when $\beta$ = 0.5 . This indicates that the introduction of contrastive learning loss does indeed help the model better focus on event pairs of the query and demonstrations, understand causalities, and achieve better performance. However, it is important to strike a balance between the contrastive learning loss and the prediction loss. Excessive emphasis on the former should be avoided as it may cause the model to overly prioritize modeling event pairs and overlook the semantic relevance of the context, which can ultimately lead to a decrease in the model's performance.

\end{document}